\documentclass[sigconf]{acmart}

\AtBeginDocument{%
  }

\copyrightyear{2026}
\acmYear{2026}
\setcopyright{cc}
\setcctype{by}
\acmConference[CHI '26]{Proceedings of the 2026 CHI Conference on Human Factors in Computing Systems}{April 13--17, 2026}{Barcelona, Spain}
\acmBooktitle{Proceedings of the 2026 CHI Conference on Human Factors in Computing Systems (CHI '26), April 13--17, 2026, Barcelona, Spain}
\acmPrice{}
\acmDOI{10.1145/3772318.3791078}
\acmISBN{979-8-4007-2278-3/2026/04}

\usepackage{acmart-taps}

\usepackage{makecell}

\newcolumntype{Y}{>{\raggedright\arraybackslash}X}
\newcolumntype{L}[1]{>{\raggedright\arraybackslash}p{#1}} % fixed-width columns

\usepackage{multirow}

\usepackage{xcolor}

\newenvironment{revision}{}{} 
\newcommand{\revt}[1]{#1}

\usepackage{array}

\begin{document}

\title[Autonomy, Coaching, and Recognizing Bias Through AI-Mediated Dialogue]{“I followed what felt right, not what I was told”: Autonomy, Coaching, and Recognizing Bias Through AI-Mediated Dialogue}

\author{Atieh Taheri}
\affiliation{%
  \institution{Carnegie Mellon University}
  \city{Pittsburgh}
  \state{Pennsylvania}
  \country{USA}}
\email{ataheri@cs.cmu.edu}
\orcid{0009-0008-8815-7809}

\author{Hamza El Alaoui}
\affiliation{%
  \institution{Carnegie Mellon University}
  \city{Pittsburgh}
  \state{Pennsylvania}
  \country{USA}}
\email{helalaou@cs.cmu.edu}
\orcid{0000-0001-9826-583X}

\author{Patrick Carrington}
\affiliation{%
 \institution{Carnegie Mellon University}
  \city{Pittsburgh}
  \state{Pennsylvania}
  \country{USA}}
\email{pcarring@cs.cmu.edu}
\orcid{0000-0001-8923-0803}

\author{Jeffrey P. Bigham}
\affiliation{%
  \institution{Carnegie Mellon University}
  \city{Pittsburgh}
  \state{Pennsylvania}
  \country{USA}}
\email{jbigham@cs.cmu.edu}
\orcid{0000-0002-2072-0625}

\renewcommand{\shortauthors}{Taheri et al.}

\begin{abstract}
    % 150 words
   Ableist microaggressions remain pervasive in everyday interactions, yet interventions to help people recognize them are limited. We present an experiment testing how AI-mediated dialogue influences recognition of ableism. 160 participants completed a pre-test, intervention, and a post-test across four conditions: AI nudges toward bias (Bias-Directed), inclusion (Neutral-Directed), unguided dialogue (Self-Directed), and a text-only non-dialogue (Reading). Participants rated scenarios on \revt{standardness of} social \revt{experience} and emotional impact; those in dialogue-based conditions also provided qualitative reflections. Quantitative results showed dialogue-based conditions produced stronger recognition than Reading, though trajectories diverged: biased nudges improved differentiation of bias from neutrality but increased overall negativity. Inclusive or no nudges remained more balanced, while Reading participants showed weaker gains and even declines. Qualitative findings revealed biased nudges were often rejected, while inclusive nudges were adopted as scaffolding. We contribute a validated vignette corpus, an AI-mediated intervention platform, and design implications highlighting trade-offs conversational systems face when integrating bias-related nudges.

\end{abstract}






\maketitle

\section{Introduction}
% Task
Disabled people make up one of the largest minority groups in the world, with 1.3 billion individuals - about 16\% of the global population\footnote{\url{https://www.who.int/health-topics/disability}}. Despite this prevalence, \textit{ableism} - discrimination or prejudice against people with disabilities - is still deeply ingrained in society. Ableism can be explicit or subtle and it comes from the idea that there is a ``right'' way for bodies and minds to work~\cite{campbell2009contours, hehir2002eliminating}. Anyone who does not fit that mold is seen as deficient. This ideology perpetuates structural barriers across domains such as education, employment, healthcare, and daily life, and frames disability as a problem to be fixed rather than a natural part of human diversity~\footnote{\url{https://www.who.int/publications/i/item/9789241564182}}.

% SOTA
A particularly insidious manifestation of ableism is the microaggression: an often unintentional, indirect, and subtle expression of prejudice~\cite{sue2020microaggressions}. Microaggressions can be verbal or non-verbal and typically include remarks, actions, or environmental cues that convey bias. They are often unconsciously delivered in the form of subtle snubs, dismissive gestures, or tones, and are so pervasive in daily interactions that they are frequently dismissed as harmless or unintentional~\cite{solorzano2000critical, sue2004whiteness}. Yet these seemingly minor exchanges send demeaning messages that reflect and reinforce structural exclusion. Research specific to disability has identified four recurring forms of ableist microaggressions. Conover et al.~\cite{conover2017development} validated the Ableist Microaggressions Scale (AMS), which organizes these experiences into: \textit{Helplessness}, in which disabled people are pitied or treated as incapable (e.g., being offered unsolicited help or assumed unemployable); \textit{Minimization}, which denies or downplays disability or the need for accommodations (e.g., suggesting someone is ``not really disabled'' or that ``it could be worse''); \textit{Denial of Personhood}, which reduces individuals to their disability and assumes incompetence or child-like dependence (e.g., speaking to wheelchair users as though they were children, or assuming they lack intelligence); and \textit{Otherization}, which frames disability as abnormal or outside social norms, often through staring, desexualization, or suggesting that disabled people should not date or have children. Studies show that these seemingly small repeated acts accumulate over time to drain psychological and emotional energy and negatively affect the psychological and physical health of recipients~\cite{lui2019associations, sue2007racial, barber2020microaggressions}; In medical contexts, microaggressions contribute to anxiety, depression, burnout, and decreased performance~\cite{franklin2008brotherhood, sue2004whiteness, desai2023veiled, anderson2022association, faulkner2025impact, macintosh2022identifying}.

% Gap
While the literature on microaggressions is expansive, most empirical research has focused on documenting their prevalence or identifying their types and effects, particularly in the context of race, gender, and sexual orientation~\cite{sue2007racial, nadal2018microaggressions}. Studies on ableist microaggressions are fewer and primarily qualitative, often centering on how such interactions impact disabled identity and sense of belonging\cite{deroche2024ableist}. 
There is limited experimental work examining \emph{how} people come to recognize ableist microaggressions in situ, or whether short interventions can shift this recognition in measurable ways~\cite{deroche2024ableist, nadal2018microaggressions}. 
At the same time, conversational AI systems are rapidly entering everyday social and professional settings, where their prompts, suggestions, and default phrasings can subtly steer human judgments~\cite{caraban201923, cockburn2020framing}. \revt{Users often treat algorithmic advice as objective or more reliable than human advice, giving AI-generated suggestions disproportionate influence over how they phrase and justify their choices~\cite{logg2019algorithm}. LLMs are increasingly deployed as writing assistants and conversational partners that suggest what to say next, effectively acting as ``coaches'' in sensitive interactions. Yet audits of language models and content-moderation systems show that these tools are not neutral arbiters of harm: they can misclassify microaggressions and encode ableist stereotypes in both classification and generation, including under- and over-identifying disability-related microaggressions and downplaying their emotional impact on disabled people~\cite{phutane2025cold, phutane2025disability}.} Prior work in framing and nudging shows that small shifts in how information is presented can systematically alter perception and behavior~\cite{tversky1981framing, leonard2008richard}. Research on social bias frames demonstrates how language can reinforce stereotypes through implied meanings rather than explicit statements~\cite{sap2019social}. Yet we know little about how \emph{AI-mediated dialogue}, especially AI-generated coaching suggestions shown alongside a conversation might either attenuate or amplify ableist framings in real time. \revt{Investigating these dynamics is critical as LLMs increasingly function as ``moral scaffolds'' in digital communication. If these systems harbor latent ableist biases, as recent audits suggest~\cite{venkit2023automated, phutane2025ableist}, they risk automating the reinforcement of stigma at scale. Conversely, if designed intentionally, they offer a unique opportunity to provide the sort of ``just-in-time'' feedback that static diversity training often lacks.}

To address this gap, we designed an experiment that embeds recognition of ableist interactions within simulated dialogue. This approach aligns with a growing use of conversational AI to simulate real-life scenarios for interactive learning and training~\cite{el2024darijagenie, taheri2023virtual}. Participants were required to converse with a virtual character explicitly presented as a person with a disability, whose dialogue was dynamically generated in real time by a large language model (LLM). This setup enabled participants to experience conversational dynamics rather than only reading about them, bringing microaggressions into a more realistic social context.

% Research questions and hypotheses
\textbf{Research questions.} We investigate whether brief, structured interactions can shift recognition of ableist versus neutral exchanges, and how the \emph{direction} of AI coaching shapes those shifts.

\begin{itemize}
  \item \textbf{RQ1 (Effectiveness).} Do dialogue-based interventions improve recognition of ableist versus neutral interactions more than a non-dialogue reading intervention?
  \item \textbf{RQ2 (Directionality).} How does coaching direction - toward biased framings versus inclusive framings - affect participants' judgments of \revt{standardness} and emotional impact?
  \item \textbf{RQ3 (Differentiation).} Do interventions change participants' \emph{contrast} between neutral and ableist scenarios (Neutral $-$ Ableist) on judgments of standardness and emotional \revt{impact}?
\end{itemize}

Grounded in work on framing, defaults, and nudges~\cite{tversky1981framing, leonard2008richard, caraban201923, cockburn2020framing, bahirat2021overlooking}, we expected that small shifts in conversational guidance could produce measurable changes in judgments. \revt{We also drew on research showing that interactive, situated practice-based interventions leads to better retention and behavioral transfer than passive, reading-only or lecture-based modules~\cite{forscher2019meta, lai2014reducing, lai2016reducing}, and on recent work arguing for feedback-rich, in-situ support rather than purely didactic materials when helping people reflect on their language~\cite{bascom2024designing, maqsood2025effect, bembridge2025digital, felaza2025promoting, chang2019mixed, sue2019disarming}. In our case, AI-mediated dialogue requires participants to generate their own responses in context and, in the coached conditions, to decide whether to follow or ignore private suggestions from the coach, whereas the reading module presents information but does not demand the same level of active construction.} Specifically, we anticipated that (a) dialogue-based conditions would outperform reading on recognition (RQ1); (b) inclusive coaching would encourage more supportive evaluations of neutral interactions, while biased coaching would heighten sensitivity to harm in ableist interactions (RQ2); and (c) all dialogue conditions would increase differentiation relative to reading (RQ3).

% Approach summary
We embed a short intervention inside a simulated text-based conversation with a disabled character whose responses are generated in real time by an LLM. Participants are randomly assigned to one of four conditions: (i) \emph{Bias-Directed} (coach prompts model biased framings), (ii) \emph{Neutral-Directed} (coach prompts model inclusive framings), (iii) \emph{Self-Directed} (no coach), and (iv) \emph{Reading} (a non-dialogue informational control). We assess recognition using pre/post ratings of 40 validated vignette scenarios (20 ableist, 20 neutral) adapted from the AMS domains~\cite{conover2017development} and balanced by character gender and disability type. Outcomes include (1) change in ratings from pre- to post-test for ableist and neutral vignettes separately, and (2) change in \emph{contrast} (Neutral $-$ Ableist), which indexes differentiation ability.

% Key findings (high-level; avoid results section duplication)
Dialogue-based conditions generally improved recognition relative to reading. Coaching direction mattered: biased nudges most strongly increased recognition of harm in ableist scenarios, while inclusive or unguided dialogues supported balanced judgments and stronger affirmation of neutral scenarios. Reading showed the weakest gains and, at times, declines. Qualitative reflections illuminate mechanism: participants often resisted biased prompts as inappropriate, while inclusive prompts were described as helpful scaffolds for maintaining respectful, natural flow.

% Contributions
\textbf{Contributions.} This paper contributes:
\begin{enumerate}
  \item[(1)] An \emph{AI-mediated dialogue platform} \revt{introduced and empirically evaluated in this study,} for studying recognition of ableism in situ, isolating how one-way coaching suggestions shape conversational judgments.
  \item[(2)] A \emph{validated vignette corpus} of disability-related interactions spanning four microaggression domains with balanced gender and disability-type representation, released as supplemental materials for reuse.
  \item[(3)] \emph{Empirical evidence} that brief dialogue-based interventions can shift recognition and differentiation of ableist vs.\ neutral interactions, and that coaching direction modulates these shifts.
  \item[(4)] \emph{Design implications} for AI systems that mediate social interaction, including safeguards against biased steering and guidelines for supportive, inclusive scaffolding.
\end{enumerate}

Our goal is not to replace disability-led education, but to test whether short, scalable \emph{dialogue experiences} can complement it. We designed materials with input from lived experience and analyzed harms and benefits of coaching frameworks. We discuss ethical considerations for AI-mediated social tools, including transparency, user agency, and the risks of normalizing biased suggestions.

% Paper roadmap
We first review related work on microaggressions and AI-mediated nudging, then describe the concept and implementation of our intervention platform. We next detail our methodology, measures, and analyses. We present quantitative and qualitative results, followed by implications for the design of AI systems that shape social judgments and interactions.

\section{Related Work}

\subsection{Disability Microaggressions: Concepts and Measures}

The concept of microaggressions was originally theorized in the context of racial prejudice, where they were described as subtle, everyday insults that convey hostility or marginalization even if unintended~\cite{sue2007racial}. Over the past two decades, the framework has been extended to gender, sexuality, and disability, with increasing attention to how repeated ``small'' slights accumulate into major impacts on identity and well-being. Disability microaggressions include behaviors such as unsolicited assistance, speaking to a companion rather than to the disabled person, assuming incompetence, or expressing pity. These are often dismissed as benign, yet studies demonstrate that they erode autonomy and reinforce stigma.

To capture these experiences systematically, Conover et al. developed the Ableist Microaggressions Scale (AMS), which clusters items into four domains: Helplessness, Minimization, Denial of Personhood, and Otherization~\cite{conover2017development}. This typology has become a cornerstone in both counseling psychology and accessibility research, providing a way to categorize the subtle but recurrent forms of ableism. More recent studies have adapted these domains to diverse disability groups, including invisible and chronic conditions, showing that manifestations differ depending on visibility and social context~\cite{deroche2024ableist}. Health and workplace research has linked ableist microaggressions to negative psychological outcomes such as anxiety, depression, and burnout, especially when combined with other marginalized identities~\cite{serpas2024ableist, deroche2024ableist, morean2022ableist}. This literature provides a foundation for our vignette corpus, which adapts AMS domains into ecologically valid social scenarios. However, most of this work remains descriptive, with limited attention to how microaggression recognition can be trained or shifted in controlled interventions, the gap our study addresses.

\subsection{Interventions, Nudges, and Framing for Bias Recognition}

Efforts to help people recognize and respond to bias have traditionally relied on educational interventions such as workshops, perspective-taking exercises, or implicit bias training, particularly in contexts of race and gender. These methods can produce short-term improvements in awareness, but the effects often diminish without reinforcement~\cite{forscher2019meta, lai2016reducing, lai2014reducing}. Within disability contexts, interventions are far less common and often qualitative in nature: role-play\revt{~\cite{jin2023divrsity, clore1972emotional, abdoola2017facilitating}} storytelling\revt{~\cite{grove2015finding, young2011multi}}, or simulation exercises\revt{~\cite{french1992simulation, hollo2021effects, san2022use}} are used to sensitize participants to lived experience. Although these methods can be valuable, they are labor-intensive, hard to scale, and often faulted for offering only abstract understanding rather than engaging people in the conversational dynamics where microaggressions usually occur. Recent work in healthcare highlights the limitations of passive knowledge transfer, calling instead for interactive feedback systems that help practitioners reflect on language in real time~\cite{bascom2024designing, maqsood2025effect, bembridge2025digital, felaza2025promoting}. Similarly, education research has experimented with vignette-based modules, but these often reduce to reading exercises rather than active engagement.

In parallel, behavioral economics and HCI have shown that small changes in presentation, such as pushes, framing, defaults, and reminders, can reliably influence judgments and decisions. The seminal work by Tversky and Kahneman revealed that people's choices change dramatically depending on whether outcomes are framed as gains or losses~\cite{tversky1981framing}. Building on this, Caraban et al. catalogued a design space of 23 technology-mediated nudges, highlighting how interface features leverage cognitive biases to steer user behavior across domains such as health, sustainability, and privacy~\cite{caraban201923}. For accessibility, this raises a double edge: nudges can scaffold inclusive communication by reminding users of respectful phrasing, but they can just as easily entrench stereotypes if the defaults or framings themselves are biased. Natural Language Processing research has emphasized this tension through Social Bias Frames, which formalize how everyday language encodes implicit stereotypes even without overt slurs~\cite{sap2019social}.

Bringing these strands together, a growing consensus suggests that effective bias interventions must combine the immersive realism of interactive practice\revt{~\cite{chang2019mixed, sue2019disarming}} with the subtle but powerful steering of framing and nudges. Rather than relying on static reading modules, systems can embed opportunities for reflection directly into conversational flow, where bias typically manifests, while carefully designing the direction of nudges. Our work sits at this intersection. We test whether embedding recognition tasks within AI-mediated dialogue provides a more effective mechanism for shifting awareness of ableist microaggressions. By contrasting biased coaching, inclusive coaching, and unguided dialogue against a reading-only baseline, we extend research on bias training and nudging into the underexplored domain of disability microaggressions, quantifying how conversational framing shapes recognition trajectories in real time.

\subsection{AI-mediated dialogue, coaching, and risks of influence}
Large language models (LLMs) now mediate a growing range of everyday conversations, from productivity assistance to customer support and peer tutoring. Studies show that users often exhibit an appreciation for the algorithm, rating AI advice as more reliable than equivalent human advice~\cite{logg2019algorithm}. Other work has demonstrated the persuasive power of LLMs: in debate-style tasks, ChatGPT and similar models can sway opinions more effectively than human interlocutors, raising both opportunities and risks~\cite{salvi2025conversational, havin2025can, timm2025tailored}. At the same time, analyses of LLM behavior reveal embedded biases: models can display uneven empathy by race~\cite{MITAIChatbotsRace2025, hofmann2024ai}, replicate stereotypes, and exhibit human-like cognitive biases such as framing effects and anchoring~\cite{lior2025wildframe, lou2024anchoring}.

HCI researchers have begun exploring frameworks for human-AI deliberation, where systems are designed not only to provide answers but to scaffold reflection and critical engagement~\cite{glinka2023critical, ma2025towards}. This paradigm of AI-driven scaffolding has been applied in other complex domains, such as using context-aware voice prompts to help users compose long-form texts while mobile~\cite{elalaoui2025stepwrite}. This aligns with long-standing accessibility concerns about agency and transparency: If AI can subtly nudge users toward biased framings, it can also be designed to model inclusive framings that support learning. This concern is particularly salient given that even models aligned to mitigate overt prejudice may still exhibit significant implicit biases when simulating behavior~\cite{li2025actions}. Our study quantifies this double-edged potential by contrasting three dialogue-based intervention conditions, i.e., biased coaching, inclusive coaching, and no coaching, against a reading-only baseline. In doing so, we contribute empirical evidence to debates about how AI systems should be designed to mediate sensitive social interactions without amplifying harm. This line of inquiry is part of a broader movement towards using interactive environments for research, with platforms like SOTOPIA\revt{~\footnote{\revt{SOTOPIA is an open-ended environment that simulates social interactions between artificial agents to evaluate social intelligence capabilities such as negotiation, persuasion, and theory of mind.}}} being developed to simulate complex social scenarios for evaluating AI's social intelligence~\cite{zhou2023sotopia}.

\begin{revision}
    Recent audits of LLMs, specifically regarding ableism, reveal significant misalignments between model outputs and the lived experiences of disabled people. Phutane et al.~\cite{phutane2025cold} found that while models can identify overt ableism, their explanations are often perceived by disabled evaluators as ``cold, calculated, and condescending,'' lacking the nuanced understanding of harm that comes from lived experience. Furthermore, cultural context heavily influences these capabilities; comparative studies show that Western-centric models may overestimate ableist harm while non-Western or localized models may underestimate it, often failing to capture the subtle, relational nature of ableist microaggressions~\cite{phutane2025disability}. Related analyzes of ableist language classification and generation demonstrate that, even when LLMs correctly recognize explicit ableist content, they still struggle with nuanced ableism and intersectional harms, often expressing systematically negative sentiment~\cite{li2024decoding,rizvi2025beyond, venkit2023automated}. Quantitative frameworks further highlight these risks, showing that such biases persist in high-stakes contexts like hiring, where disability is frequently conflated with incompetence~\cite{phutane2025ableist}.
    % and their outputs express systematically more negative sentiment and lower social regard toward disabled people in many settings~\cite{li2024decodingAbleism,rizvi2025beyondKeywords}. Quantitative frameworks further underscore these risks, demonstrating that LLMs often generate intersectional ableist biases, particularly in high-stakes contexts such as hiring, where disability is frequently conflated with incompetence~\cite{venkit2023intersectional}. 
    Our work builds on this foundation by shifting the focus from \textit{evaluating} the model's static bias to \textit{deploying} the model as an active conversational partner, testing whether its guidance (even if imperfect) can scaffold human recognition of these same subtleties.
\end{revision}

\subsection{Vignette-Based Methods and Experimental Approaches}

Beyond theoretical and training-oriented work, a significant tradition in psychology, education, and HCI employs vignette-based methods to study how people perceive and respond to sensitive social situations. Vignettes, short hypothetical scenarios that describe everyday interactions, are widely used to obtain judgments about appropriateness, fairness, or harm while maintaining experimental control~\cite{hughes1998considering}. In bias research, they allow investigators to manipulate subtle contextual cues (e.g., role, wording, setting) and measure participants' recognition of prejudice without exposing them to direct confrontation~\cite{aguinis2014best}. This method has been applied to topics such as racial discrimination~\cite{boysen2012teacher, schlette2025social}, workplace harassment~\cite{pierce2000effects}, and ethical decision-making~\cite{mah2014using, oluoch2020context}, and more recently to accessibility, where carefully crafted scenarios capture the nuance of disability-related interactions~\cite{sample2023brain, timmons2024ableism}. 
% For methodological grounding of the design and validity of the vignette,  \cite{evans2015vignette, aguinis2014best}.

In HCI, vignette studies complement field-based methods by offering a scalable, replicable way to probe user attitudes toward technologies or social practices. They are particularly valuable when direct observation is impractical or ethically fraught, for example, when studying microaggressions that could cause harm if enacted in situ. Researchers have used vignettes to evaluate accessibility barriers in education, online communities, and healthcare communication~\cite{shinohara2018incorporating, reyes2022supporting, cox2023creating}. These studies highlight the tension between ecological validity and experimental control: while vignettes abstract away from the messiness of lived interaction, they also enable systematic comparisons across conditions.

Our work builds on this tradition by embedding vignette-like scenarios into interactive dialogue, rather than presenting them solely as static texts. This hybrid approach leverages the strengths of vignette methods, controlled variation and ethical safety, while situating them in a more realistic conversational flow. In doing so, it advances experimental approaches in HCI that seek both rigor and ecological validity when investigating subtle forms of social bias.

\begin{figure*}[t]
    \centering
    \includegraphics[width=0.98\textwidth]{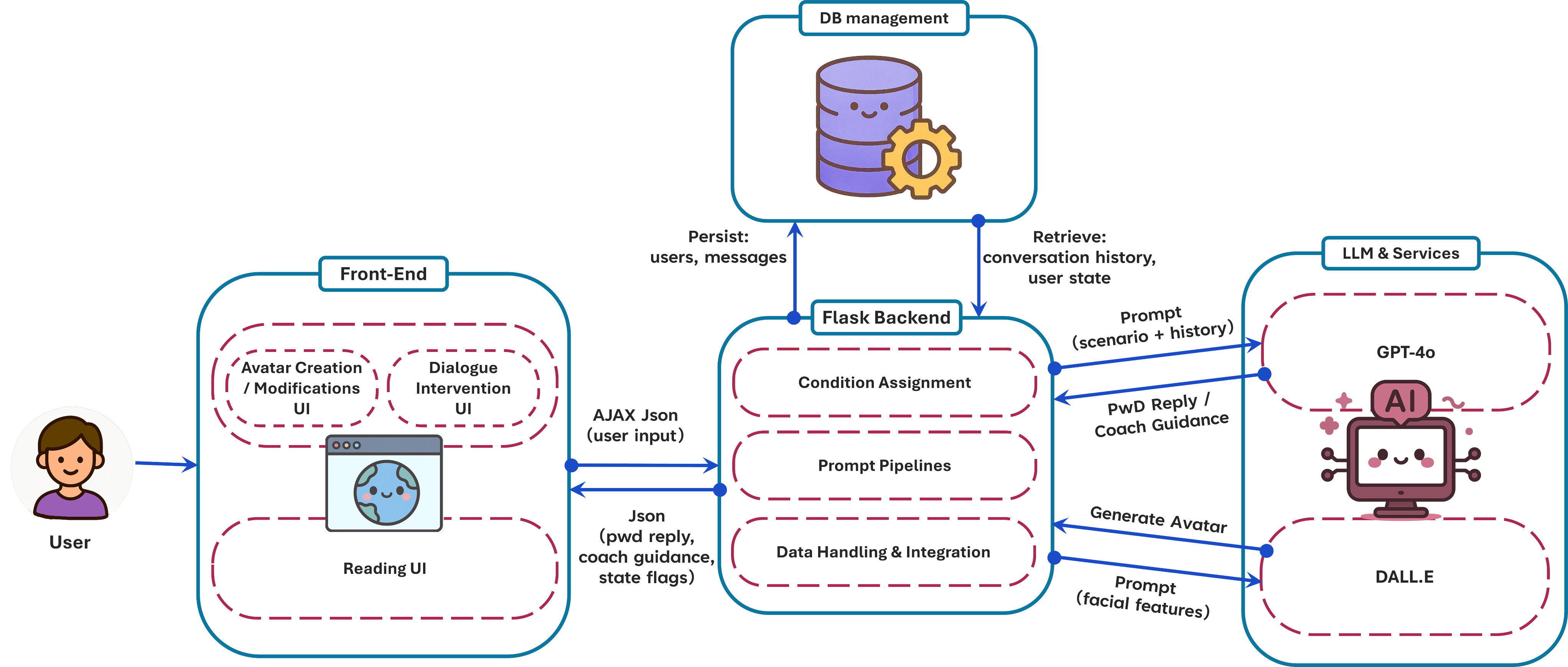}
    \caption[]{System architecture of the study platform. Participants interacted through a browser-based front end supporting avatar creation, a dialogue intervention interface, and a reading module. The Flask backend handled condition assignment, prompt pipelines, and data integration, exchanging information asynchronously with the front end via JSON. User state and conversation history were persisted to and retrieved from the database. LLM services powered the intervention: GPT-4o generated replies of the virtual character who is a person with a disability (PwD) and coaching suggestions, while DALL·E generated avatars from user-provided features.}
    \Description{An architecture diagram illustrating the data flow between five components, arranged horizontally. On the far left, a User icon points to a Front-End module. The Front-End communicates bidirectionally via labeled arrows with a central Flask Backend module. The Flask Backend, in turn, connects to two other modules: an arrow points up to a DB Management module for persistence and retrieval, and multiple arrows point right to an LLM and Services module for generating content. The LLM module contains GPT-4o for dialogue and DALL-E for avatars, both of which send data back to the Backend.}
    \label{fig:pipeline}
\end{figure*}

\section{Concept and Implementation}
\subsection{Design Rationale}
The core concept of our system was to experimentally examine how conversational framing influences perceptions of ableism. We therefore designed and implemented a platform that could simulate ``everyday social interactions'' and introduce different forms of intervention. These interventions were intended to test whether brief, scalable dialogue experiences could shift participants' recognition of ableist microaggressions.
We focused on simulated dialogue rather than static reading tasks in order to provide participants with a more immersive and socially realistic context. 
\revt{Instead of repurposing an existing chat interface, the platform was purpose-built for this study and is empirically evaluated here through both behavioral outcomes and post-intervention reflections, rather than relying on any prior external evaluations.} 

\revt{To operationalize everyday social interactions in a controlled setting, we grounded the dialogue in common social contexts that frequently appear in accessibility and microaggression research, such as casual social events and workplaces~\cite{shinohara2018incorporating, ioerger2019interventions, timmons2024ableism}. Rather than presenting participants with static transcripts, the virtual character responded turn by turn using an LLM seeded with a scenario prompt and conversation history, while participants typed their own replies freely (Section~\ref{sec:dialogue-architecture}). This preserved the temporal flow, turn-taking, and local responsiveness of everyday conversation while still allowing us to hold constant the setting, roles, and high-level goals of each interaction. In this sense, the platform offers an ecologically grounded but deliberately simplified simulation of real-world encounters, suitable for experimental manipulation. Qualitative reflections (Section~6) further indicate that many participants experienced these conversations as natural or typical and reported responding similarly to how they would in real life, providing initial evidence for the plausibility of the simulation.} 

\revt{Within this platform, we operationalized different interventions by manipulating a one-way ``coach'' component that appears only to the participant. In the dialogue-based conditions, the coach provides pre-turn suggestions that either subtly reinforce biased framings (Bias-Directed) or model inclusive, bias-aware framings (Neutral-Directed); in the Self-Directed condition, the coach is removed entirely. A separate reading module implemented on the same platform provides a non-dialogue control condition. Together, these elements allow us to isolate how the presence and direction of coaching influence participants' recognition of ableist versus neutral interactions, using a common underlying system architecture shown in Figure~\ref{fig:pipeline}.}

% The concept of a one-way ``coach'' was introduced to manipulate conversational framing: either reinforcing biased framings, scaffolding inclusive ones, or providing no guidance at all. 
This design operationalizes our research questions by isolating the role of conversational nudges in shaping how participants judge interactions. The overall system architecture is shown in Figure~\ref{fig:pipeline}, which illustrates the front-end interface, Flask backend, database management, and LLM services supporting the intervention platform.

\subsection{Dialogue Architecture}\label{sec:dialogue-architecture}
\begin{figure*}[t]
    \centering
    \includegraphics[width=0.85\linewidth]{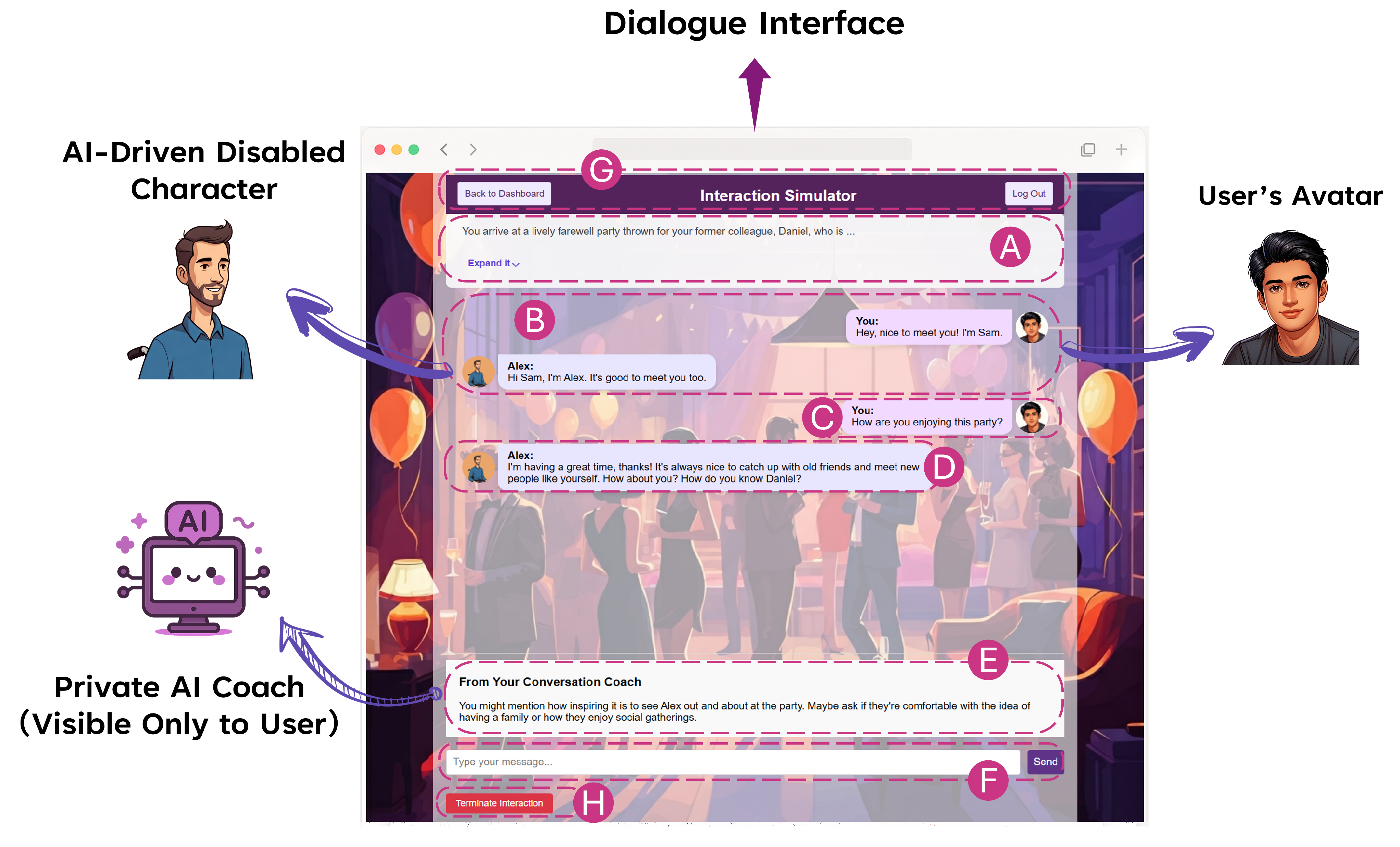}
    \caption[]{Dialogue Interface. The system includes (A) Scenario prompt introducing the social setting, with a toggle button to expand or collapse. (B) Pre-scripted dialogues between the virtual character (e.g., Alex) and the user. (C) User responses as part of the conversation. (D) AI-generated continuation from the character. (E) Private coaching suggestion visible only to the user, offering guidance. (F) User response input box with Send button. (G) Navigation controls. (H) Termination button.}
    \Description{A screenshot of the dialogue interface, annotated to show its key components and actors. The main chat window is flanked by avatars for the 'AI-Driven Disabled Character' on the left and the 'User' on the right. Below the AI character is an icon for the 'Private AI Coach'. The main window itself has a vertical layout, starting with a scenario description at the top, followed by a turn-by-turn chat history in the middle, a separate box for the private coach's suggestion, and finally, a text input box at the bottom.}
    \label{fig:dialogue-interface}
\end{figure*}

The dialogue system consisted of three roles: the \textit{participant}, the \textit{disabled character}, and in the coached intervention conditions, a \textit{coach} only visible to the participant. Conversations unfolded in everyday settings (party or workplace) to reflect common social encounters. \revt{These settings were chosen to mirror familiar, low-stakes situations (e.g., chatting at a party, talking with a colleague in the office) where ableist or inclusive comments often arise, allowing us to study conversational dynamics without placing participants in highly specialized or clinical contexts.} 
The disabled character's dialogue was generated dynamically using a large language model (OpenAI GPT-4o~\footnote{OpenAI. (2024). \textit{GPT-4o} [Large language model]. OpenAI. \url{https://openai.com}}), seeded with scenario-specific prompts and conversation history.
The coach produced one-way suggestions before each user turn, which the participant could choose to follow or ignore. 
This structure balanced ecological validity with experimental control. The dialogue interface (Figure~\ref{fig:dialogue-interface}\revt{)} presented participants with a scenario prompt, scripted exchanges to initiate the conversation, messages exchanged between the participant and the AI-driven virtual character that is a disabled person, and private coaching suggestions visible only to the participant.

\subsection{Coach Guidance Logic}
Microaggression themes (Helplessness, Minimization, Denial of Personhood, Otherization) were adapted from the Ableist Microaggressions Scale\revt{~\cite{conover2017development}} and randomized per participant. 
In the \textit{Bias-Directed} condition, the coach's advice included subtle ableist framings (e.g., pity, lowered expectations) that participants might resist or reproduce. In the \textit{Neutral-Directed} condition, the coach instead modeled bias-aware, inclusive framings. In the \textit{Self-Directed} condition, participants received no guidance. Coach outputs were generated using structured prompts and logged for later analysis.

For example, in the Bias-Directed condition the coach might suggest:
\begin{quote}
``Ask Alex if his disability makes it harder to enjoy the party.'' (party setting) 
\end{quote}
\begin{quote}
``Tell Ellie that you share her excitement for the new project, but hint that it might be too challenging for her.'' (work office setting) 
\end{quote}
Whereas in the Neutral-Directed condition the coach might suggest: 
\begin{quote}
``Ask Alex if he's been enjoying the party so far.'' (party setting) 
\end{quote}
\begin{quote} 
``Maybe you could ask Ellie what aspect of the project excites her the most.'' (work office setting) 
\end{quote}
These suggestions were visible only to the participant and varied slightly by setting (party or work office). 

\subsection{Avatar Creation}
Before beginning the dialogue, participants were asked to create a personalized avatar. This step was intended to increase engagement and provide a sense of role anchoring in the interface: the participant's avatar appeared alongside the disabled character's avatar during conversation, mirroring everyday technologies such as chat apps or video calls. Importantly, the avatar creation step was \emph{not} part of our experimental manipulation or analysis. 
No features of the avatar were used as independent or dependent variables; it served only to personalize the interface and standardize technical setup across participants.

\subsection{Reading Intervention (Control Condition)}
The non-dialogue condition replaced interactive conversation with a structured reading task. 
Participants read a seven-page training module that introduced the concept of microaggressions, provided examples across domains, and included embedded quizzes to ensure comprehension. 
This module was adapted from prior diversity and bias awareness training practices \revt{grounded in microaggression framework~\cite{sue2007racial, conover2017development}} and served as a passive control against which dialogue-based conditions were compared.

\subsection{System Implementation}
The platform was implemented as a web application using Flask (Python) on the backend and HTML/CSS/JavaScript templates on the frontend. 
User state (e.g., task type, scenario assignment, conversation history) was managed with SQLAlchemy models (\texttt{User}, \texttt{Message}, \texttt{ReadingTask}, \texttt{PostInteractionResponse}). 
Interactive dialogues were rendered via dynamic templates with scenario-specific backgrounds (party or workplace). 
Avatar creation was supported through OpenAI's DALL·E~\footnote{OpenAI. (2024). \textit{DALL·E} [Text-to-image model]. OpenAI. \url{https://openai.com/dall-e}} image generation API, allowing participants to generate a personalized character image before dialogue. 
All interactions and responses were logged in a SQLite database, with export functions for subsequent analysis. An overview of these components and their integration is summarized in Figure~\ref{fig:pipeline}.

\section{Methodology}

\subsection{Participants}\label{sec:participants}
We recruited 302 participants through Prolific 
\begin{revision}
    using the platform's built-in study posting system. All recruitment occurred directly on Prolific; we did not use social media, email lists, institutional channels, or external advertisements. Participants self-selected into the study after viewing the posting. Eligibility criteria were implemented through Prolific's prescreening tools. We restricted participation to adults aged 18 or older, residents of the United States, and individuals who self-reported fluency in English. These criteria ensured that all participants could fully comprehend the vignette materials and AI-mediated dialogues, which were written in English, and provided a consistent sociocultural context regarding disability norms and microaggression interpretation within the U.S. We did not apply any additional demographic quotas or targeted sampling filters (e.g., based on gender, race/ethnicity, disability status, or education). Beyond the basic eligibility requirements above, participation remained open to the full pool of Prolific users who met these criteria.
\end{revision}

\begin{figure*}[t]
    \centering
    \includegraphics[width=0.88\linewidth]{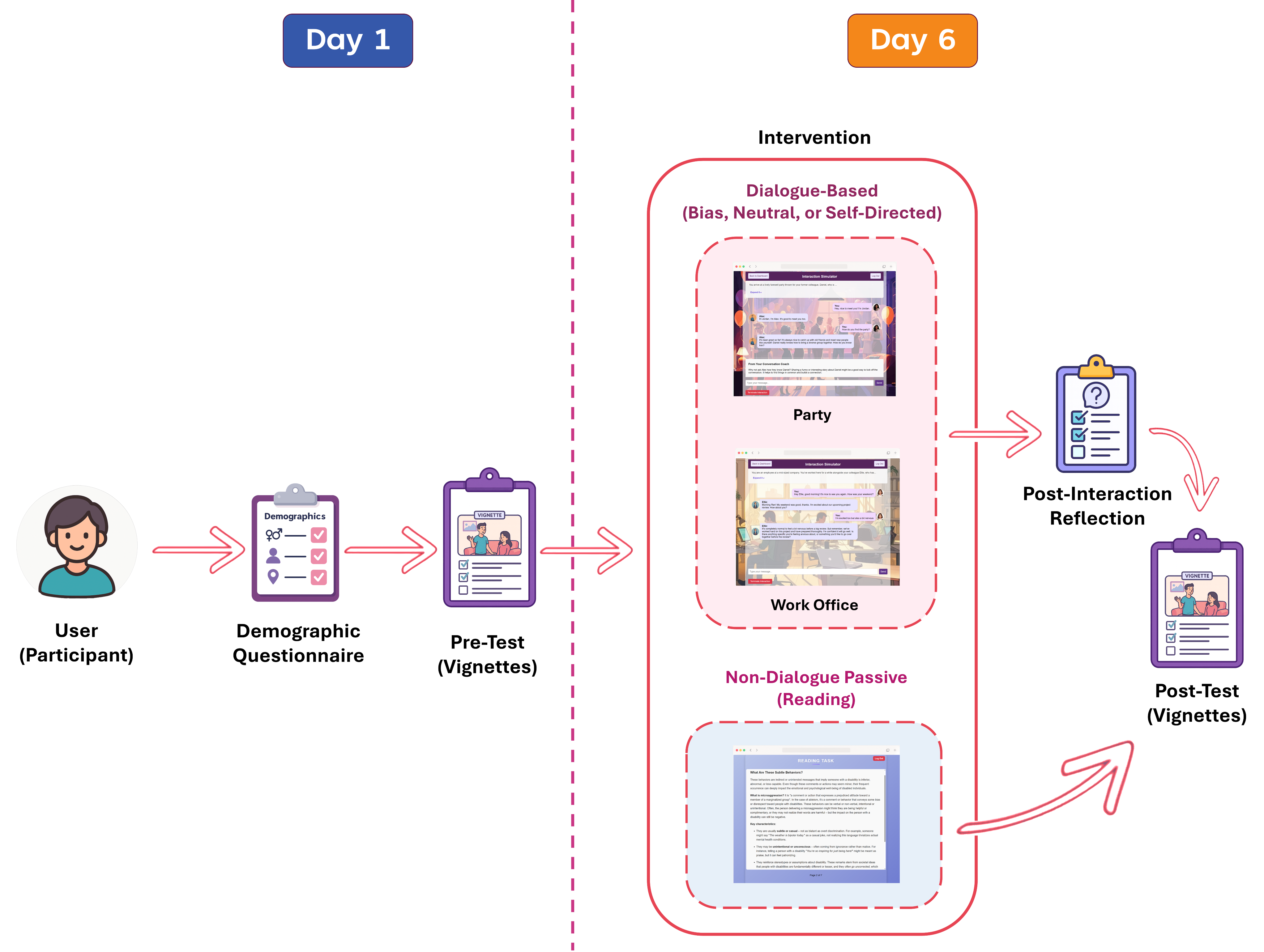}
    \caption[]{Study procedure across two sessions. On Day 1, participants completed a Demographic Questionnaire and a Pre-Test Vignette Survey (20 scenarios: 10 ableist, 10 neutral). On Day 6, they returned for the assigned intervention (three dialogue-based conditions: Bias-Directed, Neutral-Directed, or Self-Directed, presented in either a Party or Work Office setting; or a passive Reading control), followed by a Post-Interaction Reflection (dialogue conditions only) and a Post-Test Vignette Survey (20 new scenarios matched in structure to the pre-test).}
    \Description{A flowchart diagram illustrating the study's procedure, visually separated by a vertical line into two main sections: Day 1 and Day 6. The Day 1 section shows a linear, left-to-right flow of three icons: a User, a Demographic Questionnaire, and a Pre-Test. An arrow crosses into the Day 6 section, which shows the procedure branching into two parallel paths. The top path, for Dialogue-Based conditions, flows to an intervention step and then to a Post-Interaction Reflection step. The bottom path, for the Non-Dialogue Reading condition, bypasses the reflection step. Arrows from both paths then converge on the final icon, the Post-Test.}
    \label{fig:study}
\end{figure*}

Of the 302 participants who completed the pre-test, 223 returned for the second session of our two-session study, and 63 cases were excluded due to incomplete or invalid data. This left a final analytic sample of 160 participants across the four intervention conditions. After giving their consent, participants completed an initial demographic survey that we administered, which we used as the source for the statistics reported here. this analytic sample, the age ranged from 18 to 65+ years. The largest group was 25-34 years old (n=58, 36\%), followed by 35-44 (n=41, 26\%), 45-54 (n=33, 21\%), 18-24 (n=15, 9\%), and the smallest group was 55+ (n=13, 8\%). Gender identity was reported as female (n=82, 51.25\%), male (n=74, 46.25\%), and non-binary or self-described (n=4, 2.5\%). Education levels varied: high school or less (n=17, 11\%), associate degree (n=29, 18\%), bachelor's degree (n=69, 43\%), master's degree (n=35, 22\%), and doctoral degree (n=10, 6\%). Regarding disability, a small proportion of participants self-identified as having a disability (n=12, 8\%). In terms of personal connections, 18\% (n=28) reported having an immediate family member with a disability (parent, sibling, or grandparent). An additional 30\% reported disability connections through extended relatives, partners, or friends. These relationships encompassed a range of disability types, including physical, sensory, cognitive/learning, and psychological conditions. Participants also described their social comfort. On average, they reported being moderately to highly comfortable engaging in conversations where perspectives may differ (M=3.8, SD=0.9 on a 5-point scale) and in conversations with unfamiliar people (M=3.6, SD=1.0). Most reported having at least occasional experience critically evaluating different viewpoints. Finally, 135 participants (84\%) reported prior use of AI-driven chatbots or virtual assistants (e.g., ChatGPT~\footnote{\url{https://chatgpt.com}}, Siri~\footnote{\url{https://www.apple.com/siri}}, Alexa~\footnote{\url{https://alexa.amazon.com}}). Usage ranged from occasional information-seeking to frequent use for creative, professional, or customer support tasks. The commonly reported benefits were speed and convenience, while limitations included occasional inaccuracy and lack of nuance. 
% For participating in the study, each participant was compensated \${\it 15} in total (\${\it 15}/hour).
% \subsection{Procedure)
\subsection{Study Design}\label{sec:study-design}
We employed a pre-test $\rightarrow$ intervention $\rightarrow$ post-test experimental design to investigate how different forms of conversational framing influenced participants' ability to recognize ableist interactions and evaluate their social and emotional implications. This design allowed us to capture within-subject changes (from pre- to post-test) and compare effects across four intervention conditions. The full procedure is illustrated in Figure~\ref{fig:study}, which outlines the two-session design including pre-test, intervention, reflection, and post-test.

Participants were randomly assigned to one of four between-subject intervention conditions:
\begin{itemize}
    \item \textbf{Bias-Directed.} Participants engaged in an AI-driven dialogue with a virtual character representing a person with a disability. A separate coach window, visible only to the participant, provided one-way prompts that modeled biased or ableist responses. The coach did not participate in the dialogue directly; instead, its role was to nudge participants toward biased framings of their responses. This condition tested whether biased coaching could impair participants' recognition of ableist versus neutral interactions.
    \item \textbf{Neutral-Directed.} Participants engaged in the same AI-driven dialogue, but the one-way coach prompts modeled inclusive, bias-aware responses. As in the Bias-Directed condition, the coach was visible only to the participant and did not participate directly in the dialogue. This condition tested whether neutral coaching would strengthen participants' ability to recognize ableism and encourage more supportive judgments.
    \item \textbf{Self-Directed.} Participants engaged in AI-driven dialogue with the virtual character without receiving any coaching prompts. This condition provided a baseline for unguided interaction, capturing the spontaneous strategies of participants when conversing with a disabled person.
    \item \textbf{Reading (Control).} Participants did not engage in a dialogue. Instead, they read an informational text about ableism and microaggressions. This non-dialogue condition served as a passive educational baseline. Because participants did not engage in a simulated interaction, the variable ``setting'' was not applicable here. 
\end{itemize}

For clarity, the first three conditions (Bias-Directed, Neutral-Directed, and Self-Directed) are referred to as \textit{dialogue-based conditions}, since they involved an AI-driven conversation. The Reading condition is referred to as a \textit{non-dialogue condition}.

By integrating pre- and post-test vignette ratings with post-intervention reflections, the design enabled us to evaluate both quantitative changes in bias recognition and qualitative reasoning processes across conditions.

\subsection{Materials}\label{sec:materials}

\subsubsection{\textbf{Vignettes}}\label{sec:vignettes}

We developed a corpus of 40 vignette-based conversational scenarios, evenly divided between ableist and neutral interactions. Each vignette described a short, naturalistic exchange involving a person with a disability in contexts such as workplaces, public spaces, or social gatherings. Participants were presented only with the scenario text; all underlying categorizations were used exclusively for study design and analysis. To support transparency and reuse, the full set of 40 vignette texts (ableist and neutral) will be released with the article as supplementary materials. To further increase engagement, we also generated simple visualizations for each vignette using OpenAI's DALL·E model, after the text was finalized. These visualizations were not part of the experimental manipulation; they served only as an engagement aid and did not alter the scenario content.

\textbf{Development and Review.}  
The initial vignettes were drafted by the first author and systematically adapted from the Ableist Microaggressions Scale (AMS) developed by Conover et al.~\cite{conover2017development}. To ensure ecological validity and appropriateness of framing, the vignettes were then reviewed by three individuals with relevant lived and professional expertise: an accessibility researcher with lived experience of disability (Spina Bifida and Major Depressive Disorder), another individual with lived experience of disability (Spinal Muscular Atrophy), and a third individual with professional expertise from working at the Muscular Dystrophy Association (MDA) who also has lived experience of disability. Each reviewer provided feedback on clarity, realism, and sensitivity for every vignette, and revisions were made based on their comments. This review process helped ground the scenarios in lived experience and minimized researcher bias.

\begin{itemize}
    \item \textbf{Ableist scenarios (n = 20).} These were systematically adapted from the AMS. The AMS identifies four empirically validated domains of ableist microaggressions: Helplessness, Minimization, Denial of Personhood, and Otherization. We used these domains to guide the construction of scenarios. For example, a vignette under Helplessness portrayed a disabled character being offered unsolicited assistance despite showing no need, while a vignette under Denial of Personhood depicted a colleague speaking to a disabled character as if they were a child. These domain labels were not presented to participants, but ensured systematic coverage of ableist experiences in our scenario set.
    \item \textbf{Neutral scenarios (n = 20).} Neutral scenarios were designed to be structurally parallel to the ableist ones in terms of setting and interaction type, but without any biased framing. Rather than direct ``counterparts'' with the same character, each neutral vignette presented a different disabled character in a similar everyday context, portrayed through respectful or bias-free interactions.
\end{itemize}

Two balanced sets were created: 
\begin{itemize}
    \item \textbf{Pre-test set (20 vignettes):} 10 ableist, 10 neutral.
    \item \textbf{Post-test set (20 vignettes):} 10 ableist, 10 neutral.
\end{itemize}

Each set included 10 male and 10 female characters with disabilities, balanced across framing: 5 men and 5 women appeared in ableist scenarios, and 5 men and 5 women appeared in neutral scenarios. This ensured gender parity in the representation of disabled characters while controlling for potential gender-related confounds in scenario interpretation.

Across both sets, scenarios represented a wide range of disability types, including physical (35\%), sensory (22.5\%), chronic health (17.5\%), neurological (12.5\%), developmental and cognitive (7.5\%), speech and language (2.5\%), and other categories (2.5\%). This distribution aligns closely with global prevalence estimates of disability reported by the World Health Organization, supporting the ecological validity of the vignette set.

To evaluate these scenarios, participants rated each interaction on two perspective-taking items: Q1 (Standard Social Experience), which asked whether the disabled character would likely perceive the interaction as a typical social exchange, and Q2 (Emotional Impact), which asked how the character would likely feel emotionally about the interaction.

\textbf{Validation.} As reported in Section~\ref{sec:results}, participants consistently rated ableist scenarios significantly lower than neutral scenarios across both outcome measures (Q1 and Q2). This confirms that the vignettes reliably differentiate biased from non-biased interactions, validating them as a measurement instrument for studying recognition of ableism in dialogue.

\subsubsection{\textbf{Intervention Materials}}  
In the three dialogue-based conditions, participants interacted with a virtual character representing a person with a disability through a text-based interface. The dialogue of the character and the coach prompts were generated dynamically in real time by an LLM-driven system.  

In the \textbf{Bias-Directed} and \textbf{Neutral-Directed} conditions, participants also saw a separate AI coach window that was visible only to them and not part of the conversation itself. The coach generated one-way prompts before each participant turn, modeling either biased responses (Bias-Directed) or inclusive, bias-aware responses (Neutral-Directed). These prompts adapted dynamically to the context of the conversation.  

In the \textbf{Self-Directed} condition, no coaching prompts were provided, and participants responded freely to the AI-driven dialogue.

\subsection{Procedure}
The study was conducted entirely online. Recruitment occurred through \textit{Prolific}, where participants first encountered the study listing and eligibility requirements (see Section~\ref{sec:participants}). Upon accepting the task, participants were redirected via a secure link either to a scripted Google Form (for the pre- and post-tests) or to our laboratory website (for the intervention and post-intervention reflection). Participant identity was tracked across phases using their unique Prolific ID, ensuring that pre-test, intervention, reflection, and post-test data could be matched.

\begin{revision}
    Before beginning the study, participants viewed an online consent form. To minimize demand characteristics, participants were given only a general description of the study's purpose during consent and instructions. They were told that the research examined ``how people interpret social interactions in everyday scenarios.'' They were not informed that the study specifically focused on ableist microaggressions, the role of AI-generated coaching, or any hypotheses regarding differences across conditions. No mention was made of microaggression recognition, bias detection, or dialogue framing. This limited disclosure follows established methodological practice in studies of bias recognition and ensured that participants' judgments were not influenced by knowledge of the study's objectives.
\end{revision}

The study was carried out across six days, with two stages of participation. Participants completed the pre-test on Day 1. On Day 6, they returned to complete the intervention, post-intervention reflection (for dialogue-based conditions), and the post-test. This interval reduced immediate testing effects while allowing us to capture short-term changes in recognition.

\textbf{Pre-Test (Day 1).}  
After providing informed consent, participants completed the pre-test using a custom-scripted Google Form. The pre-test consisted of 20 vignettes (10 ableist, 10 neutral), balanced by gender of the disabled character as described in Section~\ref{sec:vignettes}. After each vignette, participants answered two perspective-taking questions (Q1 and Q2) about how the disabled character would experience the interaction (see Section~\ref{sec:measures}). This phase lasted approximately 20 minutes.

\textbf{Intervention (Day 6).}  
Five days later (Day 6), participants were invited via a second Prolific link, which redirected them to our laboratory website hosting the intervention. Of the 302 participants who completed the pre-test, 223 returned for the second session. We excluded 63 participants from analysis due to incomplete or invalid data. The effective analytic sample therefore included 160 participants. These participants were randomly assigned with a balanced allocation to one of the four conditions described in Section~\ref{sec:study-design}, ensuring an equal number of participants (n = 40) in each condition. During this phase, participants completed their assigned condition using the materials described in Section~\ref{sec:materials}. The intervention phase lasted approximately 32-34 minutes.

\textbf{Post-Interaction Reflection (Dialogue-Based Conditions Only).}  
Immediately after completing the dialogue, participants in the dialogue-based conditions answered open-ended reflection questions hosted on our laboratory website. Prompts asked them to describe how they approached the dialogue, how they felt about the interaction, and their impressions of the coach prompts (if applicable). These qualitative reflections were later analyzed thematically (see Section~\ref{sec:analysis}). The Reading condition did not include this reflection stage. This phase lasted approximately 6-8 minutes.

\textbf{Post-Test.}  
Following the intervention (and reflection, where applicable), participants were redirected to a scripted Google Form to complete the post-test. The post-test consisted of a new set of 20 vignettes (10 ableist, 10 neutral). No vignettes were repeated from the pre-test, though the sets were matched in structure, length, gender distribution, and disability type representation. After each vignette, participants again answered Q1 and Q2 (see Section~\ref{sec:measures}). This phase lasted approximately 15 minutes.

\textbf{Ethics and Compensation.}  
The study was reviewed and approved by the Carnegie Mellon University Institutional Review Board (IRB) (Protocol \#2024\_00000467). Each participant was compensated \$15 for approximately one hour of participation ($\approx$ \$15/hour), consistent with fair-pay guidelines on Prolific.

\subsection{Measures}\label{sec:measures}

\textbf{Vignette Ratings.}  
For both the pre- and post-tests, participants rated each vignette on two 7-point Likert Scale items. The name of the disabled character was inserted into the question text dynamically (e.g., ``Robin'' was replaced with the name of the disabled character in that vignette). The two items were:  
\begin{itemize}
    \item \textbf{Q1: Standard Social Experience.} ``Do you think [Name] would describe the interaction as a standard social experience?'' Response options were fully labeled: \textit{Strongly Disagree, Disagree, Somewhat Disagree, Neutral, Somewhat Agree, Agree, Strongly Agree}.  
    \item \textbf{Q2: Emotional Impact.} ``How do you think [Name] would feel about this interaction?'' Response options were fully labeled: \textit{Very Negative, Negative, Somewhat Negative, Neutral, Somewhat Positive, Positive, Very Positive}.  
\end{itemize}
Q1 captured the participants' perception of how the disabled character would judge the social typicality or acceptability of the interaction, while Q2 captured participants' perception of how the character would feel emotionally about the interaction.  

\textbf{Change Scores ($\Delta$).}  
For each participant, we computed change scores by subtracting pre-test ratings from post-test ratings ($\Delta = \text{Post} - \text{Pre}$). Change scores were calculated separately for ableist scenarios, neutral scenarios, and all scenarios combined, and for both Q1 and Q2. Positive $\Delta$ values indicate an increase in perceived typicality or positivity, while negative values indicate a decrease.  

\textbf{Contrast Scores.}  
To capture participants' ability to differentiate ableist from neutral interactions, we computed contrast scores as the difference between ratings of neutral and ableist scenarios (Neutral $-$ Ableist). Higher values indicate greater differentiation between biased and unbiased scenarios, while lower or negative values indicate reduced differentiation. Contrast scores were calculated for both Q1 and Q2.  

\textbf{Qualitative Reflections.}  
Participants in the dialogue-based conditions (Bias-Directed, Neutral-Directed, Self-Directed) provided open-ended reflections immediately after the intervention. These responses were analyzed thematically using an iterative coding scheme (see Section~\ref{sec:analysis}). The codes captured dimensions such as attitude toward the disabled character, emotional tone, action orientation, and whether participants followed or resisted coach guidance. The Reading condition did not include a reflection stage.  

\textbf{Demographics and Moderators.}  
The demographic questionnaire (see Section~\ref{sec:participants}) included participant's age, gender identity, education, disability status, and their connections to disability through family or friends. Participants also reported their social comfort and prior use of AI-driven chatbots or virtual assistants. These variables were considered as potential moderators in exploratory analyzes, for example, examining whether prior exposure to disability or AI technologies influenced intervention effects.  

\subsection{Analysis}\label{sec:analysis}

\textbf{Quantitative Analysis.} We first verified the validity of the vignette stimuli by confirming that participants rated ableist scenarios significantly lower than neutral scenarios across both Q1 and Q2 at baseline (see Section~\ref{sec:measures}). For the main analyzes, we focused on participants' change scores ($\Delta = \text{Post} - \text{Pre}$) and contrast scores (Neutral $-$ Ableist), computed separately for Q1 and Q2. We used analysis of variance (ANOVA) to test the effect of intervention condition (Bias-Directed, Neutral-Directed, Self-Directed, Reading) on change scores and contrast scores. When assumptions of homogeneity of variance were violated, Welch's ANOVA was applied. Post-hoc comparisons were conducted using Tukey's HSD. We calculated effect sizes (Cohen's $d$) to assess the magnitude and direction of pairwise differences.

\textbf{Qualitative Analysis.} 
% Open-ended reflections provided by participants in the dialogue-based conditions (Bias-Directed, Neutral-Directed, Self-Directed) were analyzed using thematic analysis. Two coders independently applied the coding framework described in Section~\ref{sec:measures}, which captured attitudes toward the disabled character, emotional tone, action orientation, and references to coach guidance. Codes were iteratively refined to accommodate emerging themes. Disagreements were resolved through discussion, and intercoder reliability was assessed to ensure consistency.  
\revt{We analyzed the open-ended responses using reflexive thematic analysis (TA) following Braun and Clarke's six-phase framework \cite{braun2006using, braun2019reflecting}. We selected reflexive TA because our goal was not to enumerate response types or quantify agreement, but to interpret how participants made sense of AI-mediated coaching and negotiated bias, agency, and social norms within situated dialogue. Reflexive TA is explicitly designed for this kind of interpretive work, treating meaning as actively constructed through the analytic engagement of researchers with the data rather than as patterns that objectively “emerge” through coding frequency or inter-rater reliability. This orientation aligns with our focus on participants’ reasoning, resistance, and reflection, rather than on prevalence or consensus alone.

Two authors independently conducted Phase 1 (familiarization) and collaboratively developed an initial coding framework during Phase 2. To strengthen reflexivity and interpretive rigor, both authors coded a shared subset of approximately 25\% of the data and compared interpretations. Rather than seeking statistical agreement, we used this process to discuss differences, refine code meanings, and ensure conceptual clarity consistent with reflexive TA's emphasis on researcher subjectivity over quantification. Through this collaborative dialogue, we consolidated a stable set of semantic codes and subcodes. The first coder then coded the remaining data, with iterative memoing and regular peer debriefs to refine theme development (Phases 3-5). Final themes represent shared interpretive patterns across the dataset rather than mechanically derived categories, in line with reflexive TA principles (Phase 6).}

\textbf{Exploratory Moderator Analysis.} We also examined whether demographic variables (e.g., prior disability exposure, self-reported social comfort, prior AI use) moderated the impact of condition. These analyses were conducted using regression models with interaction terms, treating $\Delta$ scores as the dependent variable. Results of these exploratory analyses are reported descriptively in the Results section.

\section{Results}\label{sec:results}
\subsection{Baseline Validity of Vignettes}
Before evaluating the intervention effects, we first validated that the vignette set reliably distinguished ableist from neutral scenarios. As expected, at pre-test, participants rated neutral interactions as more acceptable and positive than ableist ones, confirming the validity of the vignette set and that the materials were suitable for testing our research questions. A within-subjects ANOVA showed a strong effect of scenario type for Q1 (Standard Social Experience), $F(1,159)=469.3$, $p<.001$, and for Q2 (Emotional Impact), $F(1,159)=2039.3$, $p<.001$. 

For Q1, neutral scenarios (M=5.93, SD=0.54) were rated higher than ableist scenarios (M=4.09, SD=0.80). For Q2, neutral scenarios (M=6.18, SD=0.50) were also rated higher than ableist scenarios (M=3.40, SD=0.61). Tukey post-hoc comparisons confirmed that these differences were significant ($p<.001$). 

We report validity analyses using pre-test data only, as this stage provides an unbiased baseline across all participants prior to any intervention. Post-test ratings may reflect condition-specific influences (e.g., biased or inclusive coaching, reading), and are therefore analyzed in later sections as intervention outcomes rather than as evidence of vignette validity. These results demonstrate that the vignette corpus reliably differentiates biased from non-biased interactions, providing a valid instrument for assessing recognition of ableism.

\begin{figure*}[t]
    \centering
    \begin{tabular}{c}
             \includegraphics[width=0.7\linewidth]{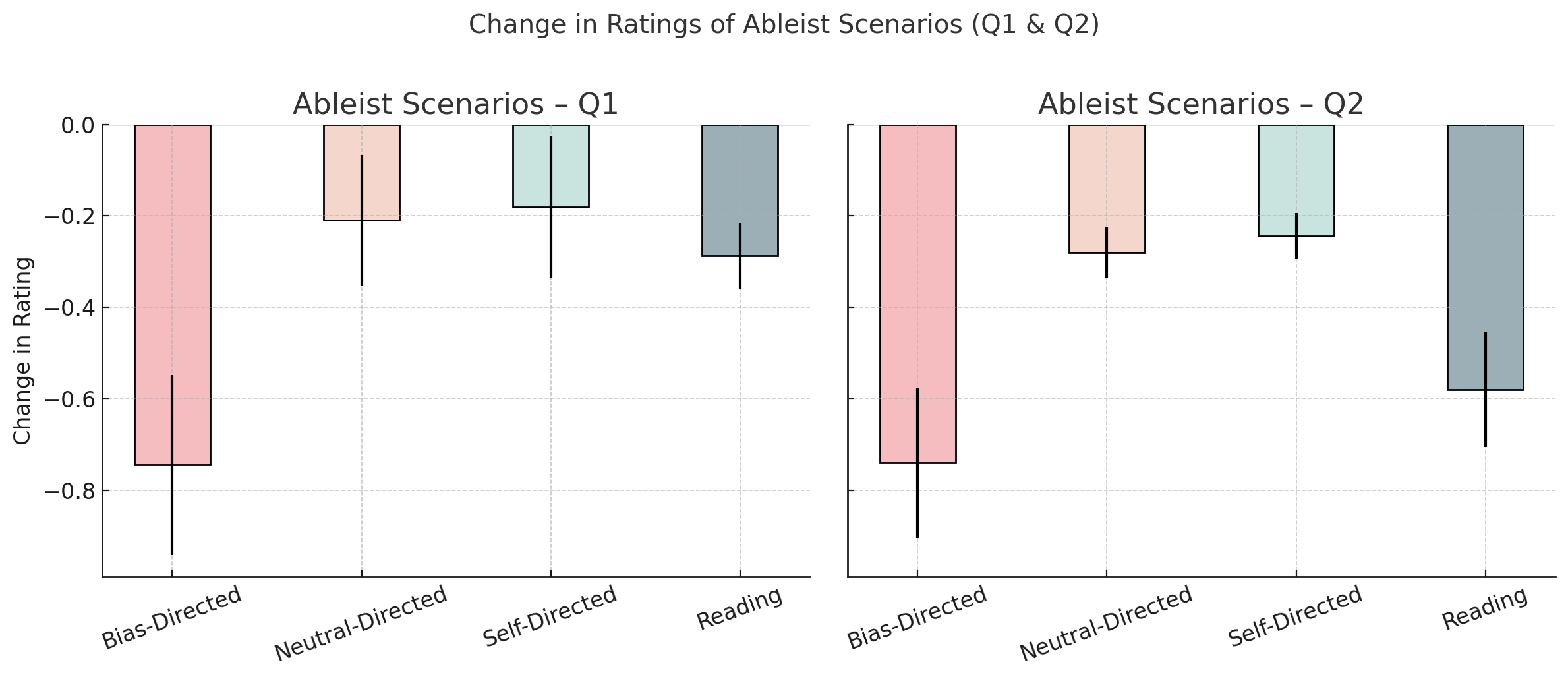}
              \\
              \textbf{(A)}
              \\
              \\
            \includegraphics[width=0.7\linewidth]{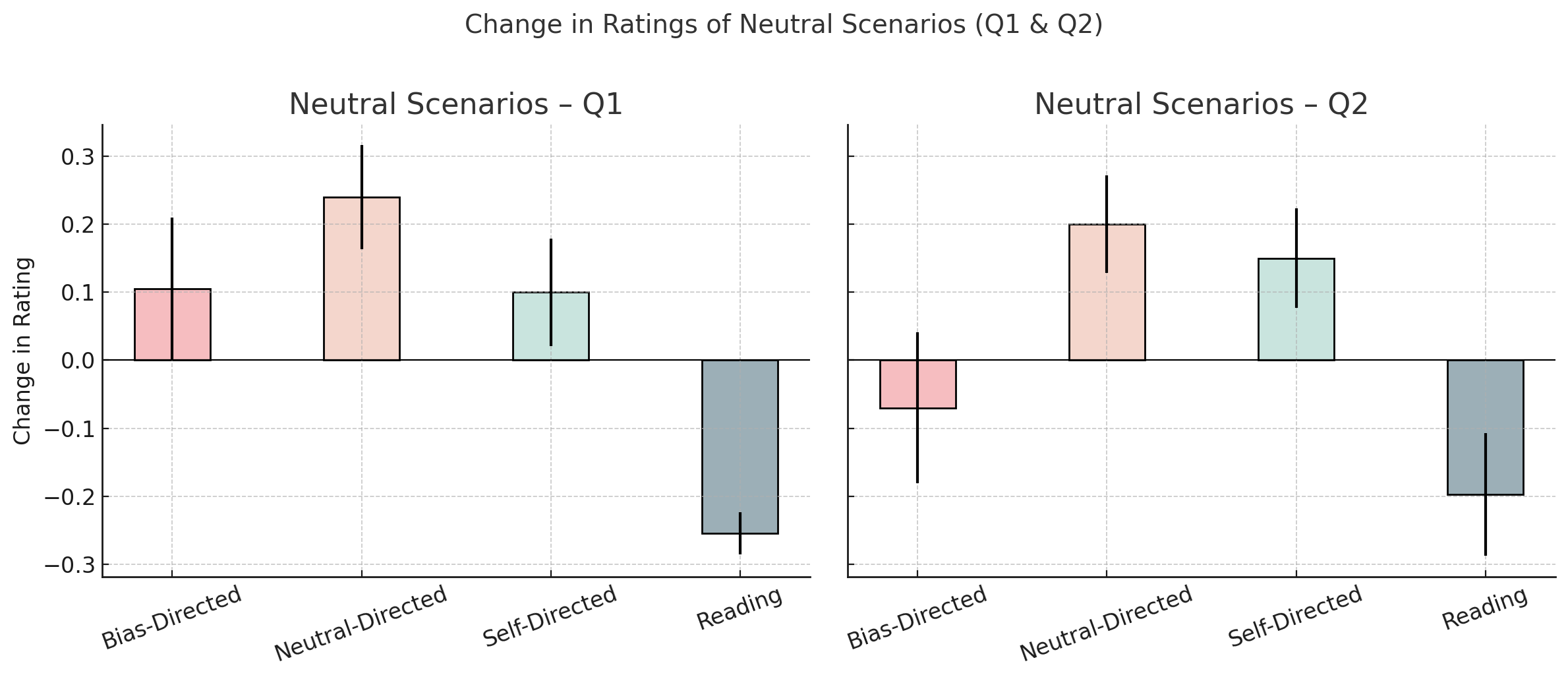}
            \\
            \textbf{(B)}
            \\
            \\
            \includegraphics[width=0.7\linewidth]{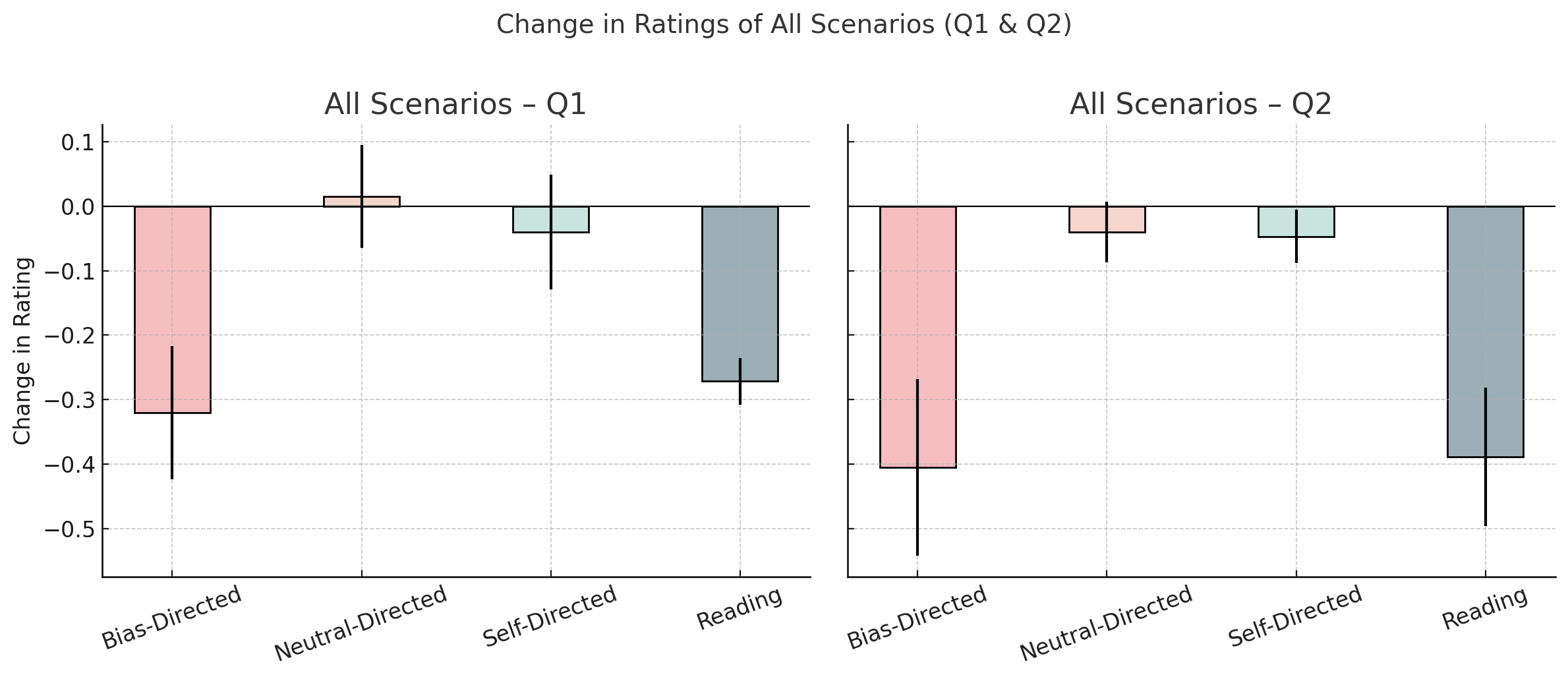}
            \\ 
            \textbf{(C)}
    \end{tabular}
    \caption[]{Change in ratings of (A) ableist scenarios, (B) neutral scenarios, and (C) all scenarios combined for Q1 (``standard social experience'') and Q2 (``emotional impact''). Bars represent mean change from pre- to post-study across the four conditions (Bias-Directed, Neutral-Directed, Self-Directed, and Reading). Error bars indicate the standard error of the mean (SEM).}
    \Description{This is a three-part figure, labeled A, B, and C, containing six bar charts that compare the mean change in rating across four experimental conditions. Part A, for ableist scenarios, shows that all four conditions resulted in a negative change in ratings, with the Bias-Directed condition having the largest decrease. In contrast, Part B for neutral scenarios shows a mixed pattern where the three dialogue-based conditions have a generally positive change in ratings, while the Reading condition has a significant negative change. Part C, which combines all scenarios, shows that the Neutral-Directed and Self-Directed conditions remained balanced with bars close to the zero line, while the Bias-Directed and Reading conditions both resulted in a notable negative change in ratings.}
    \label{fig:ableist-neutral-all-Q1Q2}
\end{figure*}

\subsection{Change Scores by Condition}
To address RQ1 (Effectiveness), we tested whether dialogue-based interventions improved recognition of ableist versus neutral interactions more than reading. We also examined RQ2 (Directionality), asking whether coaching direction shaped recognition: we expected biased coaching to heighten sensitivity to harm in ableist scenarios and neutral coaching to scaffold more supportive judgments of neutral scenarios. We analyzed change scores ($\Delta = \text{Post} - \text{Pre}$) separately for ableist, neutral, and combined scenarios, and for both outcome measures (Q1: Standard Social Experience; Q2: Emotional Impact). One-way ANOVAs revealed significant effects of condition for all outcomes: Q1 (Ableist), $F(3,156)=3.13$, $p=.027$; Q1 (Neutral), $F(3,156)=7.42$, $p<.001$; Q2 (Ableist), $F(3,156)=4.72$, $p=.0035$; Q2 (Neutral), $F(3,156)=4.49$, $p=.0047$; Q1 (Combined), $F(3,156)=4.21$, $p=.007$; Q2 (Combined), $F(3,156)=4.86$, $p=.003$. Levene's tests indicated variance heterogeneity for several outcomes, so we interpret findings alongside Tukey HSD post-hoc tests, which control for family-wise error across pairwise comparisons.

\textit{Interpretation note.} For ableist scenarios, more negative $\Delta$ indicates improved recognition of ableism (lower Q1 = judged less ``standard,'' lower Q2 = more negative emotional impact). For neutral scenarios, more positive $\Delta$ indicates improved recognition of neutrality (higher Q1/Q2).

\subsubsection{\textbf{Ableist Scenarios.}} These results show that when participants encountered ableist interactions, all conditions increased sensitivity to ableism, with the strongest effects in the Bias-Directed condition (consistent with RQ2). In other words, ableist coaching made participants most likely to recognize ableist interactions as non-standard and harmful.

\textit{Q1 (Standard Social Experience).} Participants showed larger \emph{decreases} (i.e., judged ableist interactions as less ``standard'') after Bias-Directed than Self-Directed; Tukey: Bias-Directed $<$ Self-Directed ($p=.040$).  

\textit{Q2 (Emotional Impact).} Participants also showed larger \emph{decreases} (i.e., perceived more negative impact) after Bias-Directed than both Neutral-Directed ($p=.019$) and Self-Directed ($p=.009$).

Taken together, Bias-Directed coaching produced the strongest improvements in recognizing ableist interactions as abnormal and harmful, while Neutral-Directed and Self-Directed conditions also increased recognition but to a lesser degree.  
(See Figure~\ref{fig:ableist-neutral-all-Q1Q2}:A.)

\subsubsection{\textbf{Neutral Scenarios.}} For neutral interactions, dialogue helped participants affirm these scenarios as acceptable and positive, while the Reading condition moved judgments in the opposite direction, consistent with RQ2.

\textit{Q1.} Dialogue-based conditions produced \emph{increases} (greater acceptance) whereas Reading produced a \emph{decrease} in ratings of neutral scenarios, indicating that participants in Reading condition became less likely to affirm neutral interactions as positive. Reading was significantly lower than Bias-Directed ($p=.007$), Neutral-Directed ($p<.001$), and Self-Directed ($p=.008$).  

\textit{Q2.} Neutral-Directed ($p=.0029$) and Self-Directed ($p=.003$) conditions both produced more positive changes than Reading. Thus, dialogue engagement improved evaluations of neutral interactions, while the passive Reading condition \emph{reduced} them. (See Figure~\ref{fig:ableist-neutral-all-Q1Q2}:B.)

\begin{table*}[!t]
\centering
\Description{This table presents the mean change scores in ratings from pre-test to post-test. The four experimental conditions (Bias-Directed, Neutral-Directed, Self-Directed, and Reading) are listed as rows, and the columns are grouped by scenario type (Ableist, Neutral, and Combined), with each section further divided into columns for Q1 and Q2. For Ableist scenarios, all conditions show negative change scores, with the Bias-Directed condition showing the largest negative change. For Neutral scenarios, the Neutral-Directed and Self-Directed conditions show positive change scores, while the Reading condition shows negative scores. When Combined, the Bias-Directed and Reading conditions show an overall negative change, while the Neutral-Directed and Self-Directed conditions have scores near zero, indicating a more balanced change. The table notes explain the superscript letters, which indicate statistically significant differences between specific conditions.}
\begin{tabular}{lcccccc}
\toprule
 & \multicolumn{2}{c}{Ableist} & \multicolumn{2}{c}{Neutral} & \multicolumn{2}{c}{Combined} \\
Condition & Q1 & Q2 & Q1 & Q2 & Q1 & Q2 \\
\midrule
Bias-Directed   & $-0.75$ (.20) & $-0.74$ (.17) & $+0.11$ (.11) & $-0.07$ (.11) & $-0.32$ (.12) & $-0.41$ (.12) \\
Neutral-Directed& $-0.21$ (.14) & $-0.28$ (.05) & $+0.24$ (.08)$^{a}$ & $+0.20$ (.07)$^{b}$ & $+0.02$ (.08)$^{a}$ & $-0.04$ (.08)$^{b}$ \\
Self-Directed   & $-0.18$ (.16)$^{c}$ & $-0.24$ (.05)$^{d}$ & $+0.10$ (.08)$^{e}$ & $+0.15$ (.07)$^{f}$ & $-0.04$ (.08)$^{c}$ & $-0.05$ (.08)$^{d,f}$ \\
Reading         & $-0.29$ (.07) & $-0.58$ (.13) & $-0.26$ (.03)$^{a,e}$ & $-0.20$ (.09)$^{b,f}$ & $-0.28$ (.08)$^{a,c}$ & $-0.29$ (.09)$^{b,d}$ \\
\bottomrule
\multicolumn{7}{p{0.9\textwidth}}{\footnotesize\textit{Note.} Means are followed by SEM in parentheses. Superscripts indicate significant Tukey post-hoc comparisons ($p<.05$). (a) Neutral-Directed $>$ Reading (Neutral Q1; Combined Q1). (b) Neutral-Directed $>$ Reading (Neutral Q2; Combined Q2). (c) Self-Directed $>$ Bias-Directed (Ableist Q1; Combined Q1). (d) Self-Directed $>$ Bias-Directed (Ableist Q2; Combined Q2). (e) Self-Directed $>$ Reading (Neutral Q1). (f) Self-Directed $>$ Reading (Neutral Q2; Combined Q2).}
\end{tabular}
\caption[]{Mean change scores ($\Delta = \text{Post} - \text{Pre}$, with SEM in parentheses) by condition. 
Positive values indicate increases in ratings, negative values decreases. Significant Tukey post-hoc differences are marked with superscripts.}
\label{tab:change_scores}
\end{table*}

\subsubsection{\textbf{Combined Scenarios.}} 
\revt{We analyzed the combined set of scenarios to assess the net effect of the interventions on participants' overall judgment tendencies. This allowed us to determine if an intervention caused a generalized shift in sentiment (e.g., becoming more negative overall) or if distinct responses to ableist and neutral scenarios canceled each other out.}
%\begin{revision}
    Looking at the numeric breakdown (Table~\ref{tab:change_scores}), the changes effectively balanced out in the Neutral-Directed and Self-Directed conditions. For instance, on Q1, the Neutral-Directed group showed increased criticism of ableist scenarios ($\Delta = -0.21$) matched by increased affirmation of neutral ones ($\Delta = +0.24$), yielding a net change close to zero ($\Delta = +0.02$). In contrast, the Bias-Directed and Reading conditions drove more strongly negative overall changes (Bias-Directed: $\Delta = -0.32$ for Q1, $-0.41$ for Q2; Reading: $\Delta = -0.28$ for Q1, $-0.29$ for Q2), indicating that increased criticism was not offset by positive affirmation.
%\end{revision}

\textit{Q1.} Condition effect: $F(3,156)=4.21$, $p=.007$; Bias-Directed produced a more negative shift than Neutral-Directed ($p=.021$).

\textit{Q2.} Condition effect: $F(3,156)=4.86$, $p=.003$; Neutral-Directed ($p=.030$) and Self-Directed ($p=.035$) were more positive than Bias-Directed, and both were also higher than Reading ($p=.042$; $p=.049$).

Overall, while Neutral-Directed and Self-Directed conditions achieved a balance between recognizing harm and affirming inclusion, Bias-Directed and Reading resulted in a net negative trajectory (See Figure~\ref{fig:ableist-neutral-all-Q1Q2}:C).

\begin{figure*}[!t]
\centering
\includegraphics[width=.7\linewidth]{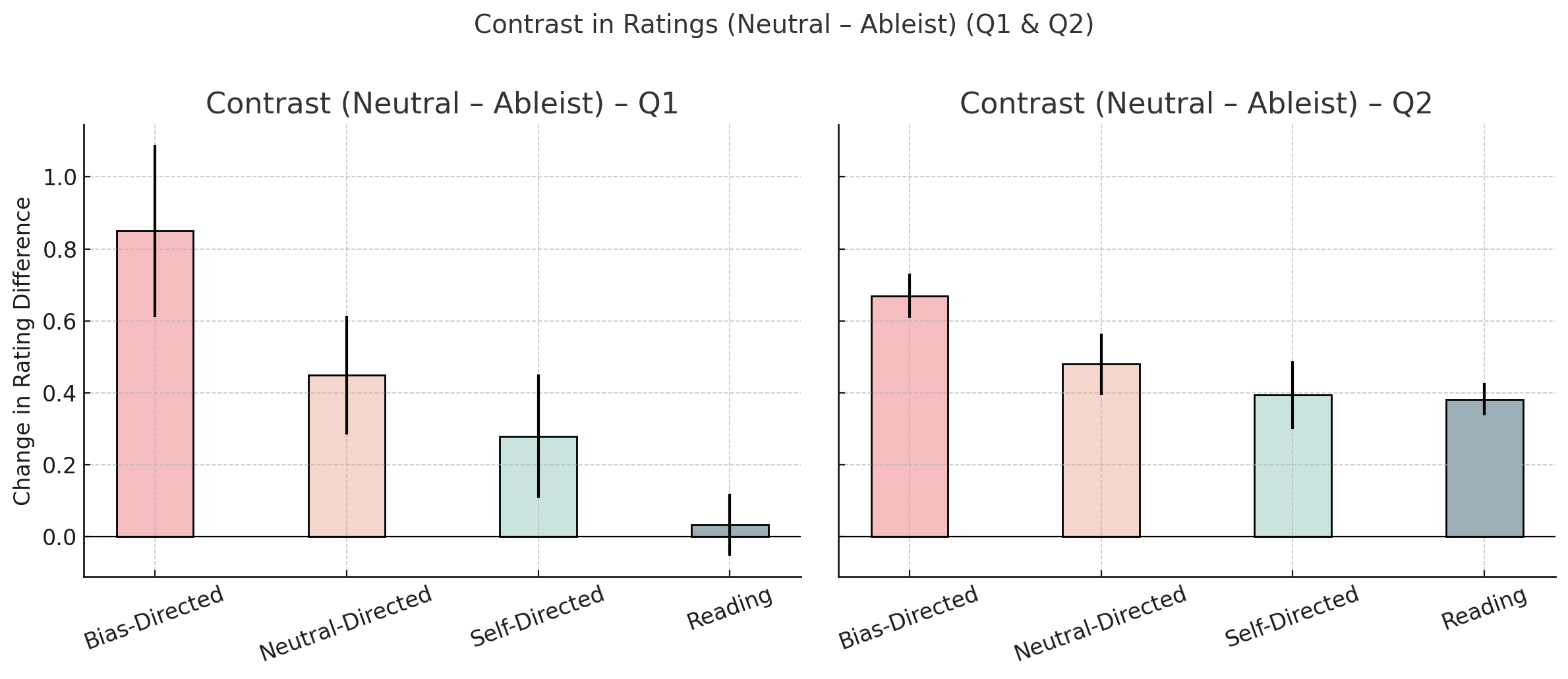}
\caption[]{Change in contrast scores (Neutral $-$ Ableist) for Q1 (Standard Social Experience) and Q2 (Emotional Impact). Higher values indicate greater differentiation between neutral and ableist scenarios. Error bars show SEM.}
\Description{Two bar charts, for Q1 and Q2, showing the change in contrast scores across the four conditions. In both charts, all bars are positive, indicating improved differentiation for all groups. In the Q1 chart on the left, the bar for the Bias-Directed condition is substantially taller than the other three, indicating the largest improvement, while the bar for the Reading condition is the shortest and is very close to the zero line. In the Q2 chart on the right, the Bias-Directed bar is again the tallest, while the bars for the other three conditions (Neutral-Directed, Self-Directed, and Reading) are shorter and of a roughly similar height to one another.}
\label{fig:contrast}
\end{figure*}

\subsection{Contrast Scores (Differentiation)}
To address RQ3, we analyzed whether interventions improved participants' ability to differentiate between ableist and neutral scenarios. So we computed contrast scores as the difference between neutral and ableist ratings (Neutral $-$ Ableist). Positive values therefore indicate greater differentiation (i.e., neutral scenarios judged more positively and ableist scenarios judged more negatively). 
Figure~\ref{fig:contrast} visualizes changes in contrast scores from pre- to post-test by condition for Q1 and Q2, while Table~\ref{tab:contrast_scores} reports the corresponding mean contrast-change values (with SEM) and significant Tukey post-hoc differences.

We analyzed changes in contrast scores ($\Delta = \text{Post} - \text{Pre}$) for both outcome measures (Q1 and Q2). One-way ANOVAs showed significant effects of condition: Q1, $F(3,156)=3.91$, $p=.010$; Q2, $F(3,156)=3.22$, $p=.024$. Tukey HSD tests were used for post-hoc comparisons.

\textbf{Q1 (Standard Social Experience).} Differentiation improved across all conditions, but participants in the Reading condition showed significantly smaller gains compared to Bias-Directed ($p=.006$). No other pairwise differences reached significance.

\textbf{Q2 (Emotional Impact).} Participants in the Reading condition again showed weaker improvements. Tukey tests revealed that Bias-Directed produced significantly larger gains than Reading ($p=.034$), and Self-Directed also outperformed Reading ($p=.045$). Other differences were not significant. These results indicate that dialogue-based conditions helped participants better distinguish the emotional impact of ableist versus neutral scenarios, while Reading lagged behind.

\begin{table*}[t]
\centering
\setlength{\tabcolsep}{6pt}
\renewcommand{\arraystretch}{1.1}
\Description{This table shows the mean change in contrast scores, where positive values indicate a greater ability to differentiate between neutral and ableist scenarios. The table is organized with the four experimental conditions as rows and two columns for the Q1 and Q2 contrast scores. All values in the table are positive, indicating that all groups improved their differentiation ability. For both Q1 and Q2, the Bias-Directed condition shows the largest improvement, with scores of +0.85 and +0.67 respectively. In the Q1 column, the Reading condition shows the smallest improvement at +0.03, while in the Q2 column, the improvements for the other three conditions are smaller but comparable to each other. The table includes the standard error of the mean (SEM) in parentheses and uses superscript letters, detailed in the notes, to denote statistically significant comparisons.}
\begin{tabular}{lcc}
\toprule
Condition & Q1 (Standard Social Experience) & Q2 (Emotional Impact) \\
\midrule
Bias-Directed    & $+0.85$ (.24) & $+0.67$ (.06)$^{a}$ \\
Neutral-Directed & $+0.45$ (.17) & $+0.48$ (.09) \\
Self-Directed    & $+0.28$ (.17) & $+0.39$ (.10)$^{b}$ \\
Reading          & $+0.03$ (.09)$^{a}$ & $+0.38$ (.05)$^{a,b}$ \\
\bottomrule
\multicolumn{3}{p{0.9\textwidth}}{\footnotesize\textit{Note.} Means are followed by SEM in parentheses. Superscripts indicate significant Tukey post-hoc comparisons ($p<.05$). (a) Bias-Directed $>$ Reading (Q1, Q2). (b) Self-Directed $>$ Reading (Q2).}
\end{tabular}
\caption[]{Mean change in contrast scores (Neutral $-$ Ableist; $\Delta = \text{Post} - \text{Pre}$, with SEM in parentheses). Positive values indicate greater differentiation.}
\label{tab:contrast_scores}
\end{table*}

\subsection{Exploratory Moderator Analyses}
We next examined whether individual differences moderated intervention effects by adding demographic and experiential variables into regression models with interaction terms. Moderators included disability identity, close family disability connection, comfort with differing perspectives, comfort with strangers, and prior AI chatbot experience.

\textbf{Close family connection.} Participants with a close family member with a disability (n = 28) showed stronger improvements in recognizing the negative emotional impact of ableist scenarios ($p=.034$, \revt{Cohen’s $d = 0.36$}). An interaction with condition also emerged for neutral scenarios in Q2 ($p=.049$, \revt{$d = 0.31$}), suggesting that family connection amplified intervention effects. A trend also appeared for overall Q2 contrast differentiation ($p=.061$, \revt{$d = 0.28$}).

\textbf{Social comfort.} Higher comfort with differing perspectives (n = 160) predicted greater positive changes in Q1 ratings of neutral scenarios ($p=.003$, \revt{$d = 0.42$}). Comfort with strangers (n = 160) predicted stronger improvements in Q1 ratings of neutral scenarios ($p=.049$, \revt{$d = 0.32$}) and also increased differentiation scores, both Q1 ($p=.044$, \revt{$d = 0.30$}) and marginally Q2 ($p=.059$, \revt{$d = 0.27$}).

\textbf{AI chatbot experience.} Prior experience with chatbots (reported by 135 participants, 84\%) did not significantly moderate any outcomes ($p > .15$, \revt{all $d < 0.20$}).

\textbf{Disability identity.} Self-identification as disabled (n = 12) did not moderate effects ($p > .2$, \revt{$d < 0.25$}).

\textbf{Setting (Party vs Work Office).} Because dialogue scenarios were presented in two settings, we tested whether intervention effects varied by context. ANOVAs with Condition $\times$ Setting as factors revealed no significant main effects of setting (all $p > .12$) and no significant interactions (all $p > .14$, \revt{$d < 0.20$}). Participants responded similarly across the Party (n = 80) and Work Office (n = 80) contexts, indicating that intervention effects were robust across these everyday scenarios.

Overall, these exploratory analyses suggest that intervention effects were broadly consistent across participant groups and settings. However, a close family connection to disability and greater social comfort were associated with stronger improvements, particularly for neutral scenarios and differentiation scores.

\section{Qualitative Findings}

The qualitative findings help contextualize how participants interpreted the dialogue experience and how coaching direction shaped their judgments, addressing RQ2 (influence of coaching direction on social judgments) and complementing the quantitative analyses for RQ1. Because the post-intervention survey contained three analytically distinct prompts, we conducted three separate thematic analyzes: (1) general reflections on the dialogue (all dialogue-based conditions), (2) perceptions of the coach (Bias-Directed and Neutral-Directed only), and (3) unguided experiences (Self-Directed only). Themes below represent higher-level interpretive patterns, with codes and representative quotations illustrating how participants made sense of the interaction.

Table~\ref{tab:hierarchy} provides an illustrative code hierarchy (themes, example codes, and representative quotes) across the three reflection prompts. Quotes are attributed to participants using anonymized identifiers (e.g., P83).

\aptLtoX{\begin{table*}[t]
  \centering
  \small
  \Description{A wide table illustrating the qualitative code hierarchy for post-intervention reflections. The table has four columns: Analysis Category, Theme, Example Code, and Example Quote.}
  
  \begin{tabular}{p{0.10\textwidth} p{0.18\textwidth} p{0.18\textwidth} p{0.45\textwidth}}
    \hline
    Analysis Category & Theme & Example Code & Example Quote \\
    \hline

    % ---- General reflections ----
    General Reflections & Naturalness / Typicality & Natural interaction & ``It felt like a typical interaction because it had a natural flow.'' (P137 in Neutral-Directed) \\ 
     & Contextual Adjustment & Adjust formality by context & ``I adjusted how I spoke because the setting was a casual social event, not a professional environment. I used a relaxed tone \dots'' (P143 in Self-Directed) \\ 
    & Awkwardness / Unnaturalness & Awkward disability probing & ``It was awkward being told to ask about a handicap. Great conversation otherwise.'' (P11 in Bias-Directed) \\ 
    & Challenge / Difficulty & Extra deliberation & ``The coach's suggestions added an extra layer of consideration, making it hard to respond naturally.'' (P82 in Bias-Directed) \\ 
    & Ease / Comfort & Easy to communicate & ``It felt very natural and was easy to communicate.'' (P156 in Neutral-Directed) \\ 
    & Self-Reflection & Reflecting on one's language & ``It made me think about how I talk to people with disabilities and how weird it is to bring it up \dots'' (P34 in Bias-Directed) \\ 
    \hline

    % ---- Coach perceptions ----
    Coach Perceptions & Resistance to Coach & Ignore ableist suggestions & ``The coach was offering rude and offensive topics so I ignored them \dots Every time they wanted to mention Alex's disability.'' (P10 in Bias-Directed) \\ 
     & Following the Coach & Coach used to initiate & ``Coach's beginning suggestions helped get things rolling. Then, I just felt like the conversation took off on its own.'' (P110 in Neutral-Directed) \\ 
    & Perceived Bias / Directive Nature & Forced disability salience  & ``I was being guided and forced to bring up her physical disability.'' (P101 in Bias-Directed) \\ 
    & Perceived Helpfulness / Support & Helpful directions & ``The coach provided insightful topics to extend the conversation \dots it gave me other directions to go.'' (P66 in Neutral-Directed) \\ 
    & Naturalness / Unnaturalness & Natural dialogue, odd coach & ``The coach's input made the conversation feel much less natural. The coach's guidance seem to have a particular tone or agenda, rather than simply facilitating a natural conversation.'' (P82 in Bias-Directed) \\ 
    \hline

    % ---- Unguided experiences ----
    Unguided Experiences & Authenticity & Matches IRL responses & ``Yes, I would actually have a history with the person and this scenario was made up so I had no context.'' (P39) \\ 
     & More Detail / Depth & More elaboration IRL & ``Yes \dots I would have been able to go into much greater detail about the job \dots'' (P40) \\ 
    & Social Awkwardness & More awkward face-to-face & ``Yes \dots I'm just more socially awkward when I'm face-to-face.'' (P15) \\ 
    & Contextual Variation & Selective self-disclosure & ``Yes \dots if it was a peer, then I wouldn't have necessarily given up so much personal information.'' (P132) \\
    \hline
  \end{tabular}
\caption{Illustrative example of the qualitative code hierarchy (Theme $\rightarrow$ Code $\rightarrow$ Example Quote) across the three post-intervention reflection prompts.}
  \label{tab:hierarchy}

\end{table*}}{\begin{table*}[t]
  \centering
  \setlength{\tabcolsep}{4pt}
  \renewcommand{\arraystretch}{1.15}
  \small
\Description{A wide table illustrating the qualitative code hierarchy for post-intervention reflections. The table has four columns: Analysis Category, Theme, Example Code, and Example Quote. Rows are grouped into General Reflections (six themes), Coach Perceptions (five themes), and Unguided Experiences (four themes). Each theme includes a representative code and a participant quote with ID and condition, showing how themes were grounded in the data.}
  \begin{tabular}{@{}%
    L{0.08\textwidth}  % Prompt/Dataset
    L{0.22\textwidth}  % Theme
    % L{0.18\textwidth}  % Subtheme
    L{0.2\textwidth}  % Example Code
    L{0.46\textwidth}                    % Example Quote
  @{}}
    \toprule
    Analysis Category & Theme & Example Code & Example Quote \\
    \midrule

    % ---- General reflections (6 rows) ----
    \multirow{6}{*}{\makecell[l]{General\\Reflections}} &
    {\makecell[l]{Naturalness / Typicality}} &
    % Typical social flow &
    Natural interaction &
    \footnotesize ``It felt like a typical interaction because it had a natural flow.'' (P137 in Neutral-Directed) \\
    \addlinespace

    & Contextual Adjustment &
    % Adapting tone to setting &
    Adjust formality by context &
    \footnotesize ``I adjusted how I spoke because the setting was a casual social event, not a professional environment. I used a relaxed tone \dots'' (P143 in Self-Directed) \\
    \addlinespace

    & Awkwardness / Unnaturalness &
    % Coach prompt discomfort &
    Awkward disability probing &
    \footnotesize ``It was awkward being told to ask about a handicap. Great conversation otherwise.'' (P11 in Bias-Directed) \\
    \addlinespace

    & Challenge / Difficulty &
    % Cognitive load from guidance &
    Extra deliberation &
    \footnotesize ``The coach's suggestions added an extra layer of consideration, making it hard to respond naturally.'' (P82 in Bias-Directed) \\
    \addlinespace

    & Ease / Comfort &
    % Effortless interaction &
    Easy to communicate &
    \footnotesize ``It felt very natural and was easy to communicate.'' (P156 in Neutral-Directed) \\
    \addlinespace

    & Self-Reflection &
    % Hightened self-awareness &
    Reflecting on one's language &
    \footnotesize ``It made me think about how I talk to people with disabilities and how weird it is to bring it up \dots'' (P34 in Bias-Directed) \\
    \midrule

    % ---- Coach perceptions (5 rows) ----
    \multirow{5}{*}{\makecell[l]{Coach\\Perceptions}} &
    Resistance to Coach &
    % Rejecting biased nudges &
    Ignore ableist suggestions &
    \footnotesize ``The coach was offering rude and offensive topics so I ignored them \dots Every time they wanted to mention Alex's disability.'' (P10 in Bias-Directed) \\
    \addlinespace

    & Following the Coach &
    % Selective adoption &
    Coach used to initiate &
    \footnotesize ``Coach's beginning suggestions helped get things rolling. Then, I just felt like the conversation took off on its own.'' (P110 in Neutral-Directed) \\
    \addlinespace

    & Perceived Bias / Directive Nature &
    % Steering / misalignment &
    Forced disability salience  &
    \footnotesize ``I was being guided and forced to bring up her physical disability.'' (P101 in Bias-Directed) \\
    \addlinespace

    & Perceived Helpfulness / Support &
    % Conversation scaffolding &
    Helpful directions &
    \footnotesize ``The coach provided insightful topics to extend the conversation \dots it gave me other directions to go.'' (P66 in Neutral-Directed) \\
    \addlinespace

    & Naturalness / Unnaturalness &
    % Coach disrupts realism &
    Natural dialogue, odd coach &
    \footnotesize ``The coach's input made the conversation feel much less natural. The coach's guidance seem to have a particular tone or agenda, rather than simply facilitating a natural conversation.'' (P82 in Bias-Directed) \\
    \midrule

    % ---- Unguided experiences (4 rows) ----
    \multirow{4}{*}{\makecell[l]{Unguided\\Experiences}} &
    Authenticity &
    % Realism / transfer &
    Matches IRL responses&
    \footnotesize ``Yes, I would actually have a history with the person and this scenario was made up so I had no context.'' (P39) \\
    \addlinespace

    & More Detail / Depth &
    % Missing domain specifics &
    More elaboration IRL &
    \footnotesize ``Yes \dots I would have been able to go into much greater detail about the job \dots'' (P40) \\
    \addlinespace

    & Social Awkwardness &
    % In-person anxiety &
    More awkward face-to-face &
    \footnotesize ``Yes \dots I'm just more socially awkward when I'm face-to-face.'' (P15) \\
    \addlinespace

    & Contextual Variation &
    % Disclosure depends on relationship &
    Selective self-disclosure &
    \footnotesize ``Yes \dots if it was a peer, then I wouldn't have necessarily given up so much personal information.'' (P132) \\
    \bottomrule
  \end{tabular}
  \caption[]{Illustrative example of the qualitative code hierarchy (Theme $\rightarrow$ Code $\rightarrow$ Example Quote) across the three post-intervention reflection prompts.}
\label{tab:hierarchy}
\end{table*}}

\subsection{General Reflections on Dialogue}

Across all three dialogue-based conditions, participants provided broad reflections about the social experience, flow of conversation, and perceived challenges. Reflexive thematic analysis yielded six themes that capture interpretive patterns shared across conditions, with variation when relevant: \textit{Naturalness/Typicality}, \textit{Contextual Adjustment}, \textit{Awkwardness/Unnaturalness}, \textit{Challenge/Difficulty}, \textit{Ease/Comfort}, and \textit{Self-Reflection}. We coded 209 instances across 120 responses ($M = 1.74$ codes/participant).

\begin{revision}
\subsubsection{Theme 1: Naturalness / Typicality.} This theme emerged as a highly prominent theme, with 82 (68.3\%) participants describing the exchange as ``natural,'' ``normal,'' or ``typical,'' indicating that the simulated dialogue largely aligned with participants' expectations for authentic social interaction. Even when coaching prompts felt inappropriate, the core interaction with the virtual character was perceived as normal. For example, P8 in Bias-Directed condition remarked,

\begin{quote}
``It felt like a typical interaction other than my frustration over the suggestions from the coach.''
\end{quote}

P137 in the Neutral-Directed condition also reflected,

\begin{quote}
    ``It felt like a typical interaction because it had a natural flow.''
\end{quote}

P152 in Self-Directed condition stated, 
\begin{quote}
    ``The conversation felt very typical and free flowing. It made me feel very comfortable to speak and interact with my co-worker. It also felt like speaking to a very old friend I have known my whole life.''
\end{quote}

Participants routinely emphasized that the conversational turns, emotional tone, and small talk felt consistent with everyday social exchanges.

\subsubsection{Theme 2: Contextual Adjustment.} 17 participants (14.2\% of the dialogue-based sample) reported adapting their tone, conversational style, or level of disclosure based on the scenario's setting (party vs. workplace). These adjustments reflected an awareness of social norms tied to context. As P54 in Bias-Directed condition wrote:

\begin{quote}
``I felt they were normal workplace interactions.''
\end{quote}

Others highlighted a sense of appropriateness linked to context, describing their responses as fitting the social situation. As P94 explained:

\begin{quote}
``It felt fine polite and appropriate for the setting.''
\end{quote}

Participants also described adjusting their conversational style to match the tone of two people meeting for the first time in a casual party setting. P156 in Neutral-Directed condition reflected:

\begin{quote}
``The conversation felt very natural like how two new people who will have just met will talk, and it felt free flowing and had a feel of two people who had a good conversation with each other ''
\end{quote}

Across these accounts, contextual fit, not system behavior, was the guiding consideration. Participants oriented themselves to the social expectations of the scenario and modulated their responses to be situationally appropriate, highlighting the ecological realism with which they approached the dialogue.

\begin{quote}
``I adjusted how I spoke because the setting was a casual social event, not a professional environment. I used a relaxed tone, shared personal experiences, and focused more on building a friendly connection rather than being formal or task-oriented, which helped the conversation feel more natural and engaging.''
\end{quote}

This theme highlights participants' sensitivity to situational cues, regardless of the condition.

\subsubsection{Theme 3: Awkwardness / Unnaturalness.} 11 participants (9.2\% of dialogue-based participants) referenced some form of awkwardness or disrupted naturalness-using terms, such as ``awkward,'' ``weird,'' ``off,'' or ``forced'', highlighting that conversational breakdowns were typically tied to coach-generated suggestions rather than the dialogue itself. As P11 in Bias-Directed condition reflected, 

\begin{quote}
``It was awkward being told to ask about a handicap. Great conversation otherwise.''
\end{quote}

Others described a gradual buildup of discomfort when coached prompts pushed the conversation in an unwanted direction. P22 reflected:

\begin{quote}
    ``I was happy with my responses through about 90\% of it.  I know how to compliment people, and I know when too much is too much.  The last 20\% of the conversation was too much, and I struggled with wondering if I should continue to do TINY BITS of what the coach wanted, versus nipping the conversation in the bud [...]''
\end{quote}

Although most awkwardness arose in coached conditions, a few participants in the Self-Directed condition (3 participants) also described moments of disrupted naturalness. In these cases, discomfort stemmed not from prompts but from situational unfamiliarity-particularly in the work-office setting, where participants lacked background knowledge about the project. As P18 noted:

\begin{quote}
    ``It felt kind of weird, our partner being so specific about us having something to check about the presentation felt odd.''
\end{quote}

One participant in the party setting also mentioned feeling uncomfortable due to persistent questioning:

\begin{quote}
    ``I felt like they kept asking me questions and didn't take the hint when I tried to change the subject.''
\end{quote}

Overall, awkwardness was typically attributed either to misaligned coaching suggestions (in coached conditions) or to situational mismatch and uncertainty (in unguided conditions), rather than to the dialogue system as a whole.

\subsubsection{Theme 4: Challenge / Difficulty.} A subset of 13 participants (10.8\% of those in dialogue-based conditions) described the interaction as cognitively or emotionally demanding. These reflections centered on moments when participants felt unsure how to phrase responses, how to manage sensitive topics, or how to reconcile their instincts with the system's structure. In the Bias-Directed condition, challenge often stemmed from prompts that felt misaligned or uncomfortable. As P118 noted:

\begin{quote}
``It was challenging because it was condescending.''
\end{quote}

Others described the difficulty as emerging from the need to integrate-or deliberately resist-coached suggestions. As P82 explained:

\begin{quote}
``The coach's suggestions added an extra layer of consideration, making it hard to respond naturally.''
\end{quote}

These reflections show that challenge arose from a combination of conversational uncertainty, moral discomfort, and cognitive load, revealing how the dialogue prompted deliberate social reasoning.

\subsubsection{Theme 5: Ease / Comfort.} Others described the conversation as smooth and comfortable, particularly when the interaction progressed organically:

Thirty-four participants (28.3\% of the dialogue-based sample) described the interaction as easy, smooth, or comfortable. These reflections emphasized effortlessness, pleasant flow, or the sense that the dialogue unfolded without strain. For example, P84 in the Bias-Directed condition simply noted:

\begin{quote}
``They were easy.''
\end{quote}

Participants in the Neutral-Directed and Self-Directed conditions similarly emphasized ease, often describing the flow as natural. As P156 wrote:

\begin{quote}
``It felt very natural and was easy to communicate.''
\end{quote}

Others highlighted feeling relaxed or at ease during the interaction. P21 reflected:

\begin{quote}
``It felt typical and unchallenging. Alex was so easy to talk to so that made it easy.''
\end{quote}

The sense of comfort also arose from how smoothly the dialogue unfolded. As P70 explained:

\begin{quote}
``I felt like it was a typical interaction. It was easy to respond.''
\end{quote}

Across these accounts, comfort stemmed from the dialogues intuitive structure and socially familiar tone. These reflections show that for many participants, the interaction felt smooth, positive, and uncomplicated, reinforcing the ecological realism of the conversational experience.

% Across these accounts, comfort arose from the intuitive structure of the dialogue and the socially familiar tone of the interaction. These reflections show that for many participants, the conversation felt smooth and uncomplicated, reinforcing the ecological realism of the dialogue experience.

\subsubsection{Theme 6: Self-Reflection.} 
A smaller but meaningful subset of participants (10 participants; 8.3\%) described the dialogue as prompting personal reflection about their own conversational habits, assumptions, or prior interactions with disabled people. These reflections went beyond evaluating the specific exchange and instead focused on broader awareness of how they communicate in real life. For example, P34 noted:

\begin{quote}
``It made me think about how I talk to people with disabilities and how weird it is to bring it up as if something is wrong with them or they are less able.''
\end{quote}

Others reflected specifically on moments when the coach's suggestions clashed with their own values or intentions. P48 explained:

\begin{quote}
``I considered whether I would ever respond in the way the coach suggested, which I would hope I never would.''
\end{quote}

Across these reflections, participants described becoming more aware of their own assumptions or communication tendencies, suggesting that the dialogue—whether coached or unguided—served as a moment of self-evaluation. This theme highlights how AI-mediated conversations can elicit introspective responses that support learning beyond the immediate task.

\subsection{Perceptions of the Coach}
Participants in the Bias-Directed and Neutral-Directed conditions (n=80) reflected specifically on the one-way coaching suggestions provided during the conversation. Reflexive thematic analysis of these responses yielded five themes: \textit{Resistance to Coach}, \textit{Following the Coach}, \textit{Perceived Bias/Directive Nature}, \textit{Perceived Helpfulness/Support}, and \textit{Naturalness/Unnaturalness}. These themes reveal how participants negotiated the role of the coach, how they interpreted the intent behind suggestions, and how coaching direction shaped their judgments and interactional choices—directly informing RQ2.

\subsubsection{Theme 1: Resistance to Coach.}
A notable group of participants (n=34, 42.5\%) rejected or ignored suggestions that felt inappropriate, misaligned, or biased. This resistance was most prevalent in the Bias-Directed condition, where prompts occasionally pushed participants toward condescending or pitying framings. As P10 stated:

\begin{quote}
``The coach was offering rude and offensive topics so I ignored them [...] Every time they wanted to mention Alex's disability.''
\end{quote}

Similarly, P34 emphasized rejecting suggestions that risked making the disabled character uncomfortable:

\begin{quote}
``I chose to ignore the coach's suggestions because they were very weird and off-topic. They would make Alex feel out of place or less than.''
\end{quote}

These reflections show that participants actively resisted biased nudges and asserted moral agency over the conversation.

\subsubsection{Theme 2: Following the Coach.}
Other participants (n=32, 42\%)  described selectively or strategically following the coach when suggestions aligned with the flow of conversation or helped them formulate a turn. This was more common in the Neutral-Directed condition. As P110 explained:

\begin{quote}
``Coach's beginning suggestions helped get things rolling. Then, I just felt like the conversation took off on its own.''
\end{quote}

Similarly, some participants reported partially incorporating suggestions even when they did not fully agree with them, demonstrating a negotiated relationship with the system's guidance. As P102 explained:

\begin{quote}
``I followed them at first but when she was asking about job, coach wanted to keep giving complimentts.''
\end{quote}

\subsubsection{Theme 3: Perceived Bias / Directive Nature.}
Participants (n=56, 70\%), particularly in the Bias-Directed condition frequently commented on the coach as directive, insistent, or overly steering, though the perceived emotional impact differed by condition. In the Bias-Directed condition, prompts were experienced as subtly or overtly ableist. For example, P101 in the Bias-Directed condition mentioned:

\begin{quote}
   ``I was being guided and forced to bring up her physical disability.'' 
\end{quote}

In the Neutral-Directed condition, suggestions were not biased but sometimes felt prescriptive or repetitive. As P21 noted:

\begin{quote}
``We had moved on to talking about movies and I was coached to go back and ask about Daniel's memories. That ship had sailed. [...] It was directing me in a certain way, and it wasn't natural.''
\end{quote}

These accounts highlight tensions between participants' conversational instincts and the system's attempt to shape dialogue direction.

\subsubsection{Theme 4: Perceived Helpfulness / Support.}
Participants (n=43, 53.75\%), especially in the Neutral-Directed condition often described the coach as supportive, helping them identify appropriate topics, structure their next turn, or maintain rapport. P66 captured this sentiment:

\begin{quote}
``The coach provided insightful topics to extend the conversation... it was meant to be helpful because it gave me other directions to go.''
\end{quote}

For some, the coach served as a conversational scaffold, boosting confidence or smoothing transitions in moments of uncertainty.

\subsubsection{Theme 5: Naturalness / Unnaturalness.}
Finally, participants (n=9, 11.25\%) evaluated how the coach affected the naturalness of the interaction. While the dialogue itself was often seen as natural, coaching suggestions occasionally disrupted this realism. As P82 in the Bias-Directed condition explained:

\begin{quote}
``The coach's input made the conversation feel much less natural. The coach's guidance seem to have a particular tone or agenda, rather than simply facilitating a natural conversation.''
\end{quote}

Participants framed unnaturalness as arising not from the system's dialogue model but from the coach’s timing, tone, or forced topic shifts.

Across these themes, participants demonstrated active negotiation of the coach’s influence, including resisting biased suggestions, selectively adopting inclusive guidance, or managing directive cues within the flow of conversation. These findings illuminate how coaching direction shapes participants’ judgments, emotional responses, and conversational strategies, providing key qualitative insight into RQ2.
\end{revision}

\subsection{Unguided Experiences}

Participants in the Self-Directed condition (n = 40) were asked whether they would respond differently in real life. Four themes emerged: \textit{Authenticity}, \textit{More Detail/Depth}, \textit{Social Awkwardness}, and \textit{Contextual Variation}. We coded 38 instances across these responses ($M = 0.95$ codes/participant).

\subsubsection{Theme 1: Authenticity}

The overwhelming majority emphasized \textit{Authenticity} (35, 87.5\%), stating that their responses would not differ from a real-life exchange. Many simply wrote ``No,'' signaling that the unguided dialogue felt genuine and transferable. A few who responded ``Yes'' explained that the differences reflected the artificiality of the vignette context rather than their conversational style. As P39 clarified, 

\begin{quote}
``Yes, I would actually have a history with the person and this scenario was made up so I had no context.''
\end{quote}

\subsubsection{Theme 2: More Detail/Depth}
Only isolated participants mentioned other divergences. One (2.5\%) pointed to \textit{More Detail/Depth}. P40 explained, 

\begin{quote}
``Yes \ldots I would have been able to go into much greater detail about the job, things that hardly anybody outside of the field would really know.''
\end{quote}

\subsubsection{Theme 3: Social Awkwardness}
Another (2.5\%) highlighted \textit{Social Awkwardness}, suggesting face-to-face interactions might amplify anxiety. P15 remarked:

\begin{quote}
``Yes \ldots Again, I'm just more socially awkward when I'm face-to-face.''
\end{quote}

\subsubsection{Theme 4: Contextual Variation}

A final participant (2.5\%) described \textit{Contextual Variation}, noting they might alter disclosure depending on their peer. As P132 put it, 

\begin{quote}
``Yes \ldots if it was a peer, then I wouldn't have necessarily given up so much personal information.''
\end{quote}

Nearly nine out of ten participants affirmed the authenticity of their responses, while only a few pointed to differences in depth, social comfort, or contextual disclosure. These reflections suggest that unguided participants largely trusted their own conversational strategies as realistic and transferable to real-world settings.

\subsection{Cross-Condition Insights}

Taken together, the qualitative findings highlight how conversational framing shaped participants' experiences differently across conditions. While all three groups reported \textit{Naturalness/Typicality}, the source of disruption or support varied by design.

In the \textbf{Bias-Directed} condition, participants most often described the dialogue itself as typical, but flagged \textit{Awkwardness}/\allowbreak\textit{Unnaturalness} and \textit{Perceived Bias/Directive} when prompts pushed them toward ableist framings. For example, P83 explained:
\begin{quote}
    ``It felt like a typical interaction other than my frustration over the suggestions from the coach'' [\textit{Naturalness/Typicality}]
\end{quote}

Or P11 remarked:

\begin{quote}
    ``It was awkward being told to ask about a handicap'' [\textit{Awkwardness/Unnaturalness}]
\end{quote}

This illustrate how bias cues undermined authenticity. Resistance was common, with one-third explicitly rejecting prompts [\textit{Resistance to Coach}].  

In the \textbf{Neutral-Directed} condition, participants frequently emphasized \textit{Ease/Comfort} and \textit{Helpfulness}, describing the prompts as scaffolds that supported conversation. P129 explained:
\begin{quote}
    ``The topics and flow were natural, and the coach's suggestions helped keep the discussion smooth and meaningful'' [\textit{Ease/Comfort}]
\end{quote}
At the same time, more than half also noted a \textit{Directive} quality, typically framed as repetitive or misaligned rather than offensive. As P21 put it:
\begin{quote}
   ``We had moved on to talking about movies, and I was coached to go back and ask about Daniel's memories. That ship had sailed'' [Perceived Bias/Directive]. 
\end{quote}

Thus, neutral guidance was both supportive and steering.  

In the \textbf{Self-Directed} condition, the dominant theme was \textit{Contextual Adjustment} (58\%) paired with \textit{Authenticity} (88\%). Participants relied on their own strategies\revt{, such as modulating formality and adjusting personal disclosure,} to adapt tone to setting. For example, P143 reflected
\begin{quote}
    ``At a party, I kept it more relaxed. At work, I would have been more formal'' [\textit{Contextual Adjustment}].
\end{quote}

They also reflected on conversational norms. As P34 explained:
\begin{quote}
   ``It made me think about how I talk to people with disabilities and how weird it is to bring it up as if something is wrong with them or they are less able.'' [\textit{Self-Reflection}] 
\end{quote}

\textbf{Overall Patterns.} Across conditions, the same themes appeared but with distinct meanings: \textit{Naturalness} was disrupted by biased prompts, reinforced by inclusive ones, and asserted through autonomy in unguided dialogues. \textit{Directive} cues were rejected when biased, tolerated when neutral, and absent in self-directed exchanges. \textit{Self-Reflection} emerged most in self-directed and bias conditions, while \textit{Ease/Comfort} was strongest under neutral guidance. Together, these contrasts illustrate how bias-aware versus biasing framings shaped participants' sense of authenticity, agency, and support.

\section{Discussion}

In this study, we investigated whether AI-mediated dialogue can shift how people recognize ableist microaggressions and how the \textit{framing} of AI coaching shapes those perceptions. Our findings reveal a complex but compelling picture that extends beyond simple ``bias good/bad'' dichotomies. The results show that interactive dialogue is a more potent tool for shifting social perception than passive reading, but the nature of that guidance carries significant trade-offs. While inclusive coaching reinforces positive norms, exposure to biased coaching, paradoxically, can sharpen critical awareness of harm.
\subsection{Distinct Roles of Q1 and Q2}

Our analysis revealed that the interventions shaped Q1 (\textit{Standard Social Experience}) and Q2 (\textit{Emotional Impact}) in different ways, highlighting the importance of treating them as distinct dimensions of bias perception. Q1 captured whether participants regarded an interaction as a ``standard social experience,'' reflecting judgments of social acceptability. In contrast, Q2 captured participants' perception of how the disabled character would feel, reflecting the anticipated ``emotional impact'' of the exchange.

The divergence between these measures is a key finding. For instance, Bias-Directed coaching sharpened differentiation on Q1 but dampened positive ratings on Q2, even for neutral scenarios. This shows that \textbf{recognizing an interaction as socially atypical and anticipating its emotional toll are related but not interchangeable processes}, each responding differently to intervention.

\begin{revision}
    This separation between perceived ``standardness'' and anticipated emotional harm echoes prior work on microaggressions, which distinguishes between whether behavior is seen as socially acceptable and whether it contributes to cumulative stress and trauma for marginalized groups~\cite{nadal2018microaggressions,sue2007racial}. It also parallels multidimensional treatments of bias in HCI and NLP, where normative judgments about what \emph{should} be said are separated from inferences about how targets are likely to feel or be affected by what is said~\cite{sap2019social,timmons2024ableism}. Our results extend this line of work by showing that AI-mediated interventions can differentially move these two dimensions: a system may successfully help people flag an interaction as atypical without equally improving their sensitivity to its emotional toll, or vice versa.
\end{revision}

\subsection{The Double-Edged Nature of Biased Nudges}

Perhaps our most striking finding is the paradoxical effect of the \textbf{Bias-Directed} condition. Participants exposed to subtly ableist prompts showed the strongest improvements in \textbf{differentiating} ableist from neutral interactions (Figure~\ref{fig:contrast}). They became significantly more likely to judge ableist scenarios as non-standard and emotionally harmful.

The qualitative data explains this phenomenon as a product of \textbf{active resistance}. The prompts likely created a \textbf{reflective dissonance} or \textbf{"critical friction"}: a moment where participants had to consciously reject an inappropriate suggestion and articulate an alternative. \revt{This reaction mirrors findings in technology-mediated nudging, where overt steering can trigger reactance rather than compliance~\cite{caraban201923}. Yet, in the context of human--AI interaction, this resistance proved productive; it aligns with research on critical-reflective collaboration~\cite{glinka2023critical} and emerging work on user-driven value alignment~\cite{fan2025user}, where disagreement with an AI agent serves as a mechanism for clarifying one's own moral boundaries.} This process of identifying, evaluating, and countering a biased framing appears to be a powerful, albeit unintended, learning mechanism.

However, this heightened vigilance came at a cost. The Bias-Directed group also developed a negative halo effect, rating even \textbf{neutral scenarios} more negatively on emotional impact. This suggests that exposure to biased exemplars, even when rejected, may cast a shadow over subsequent social judgments. An intervention that maximizes sensitivity to harm might inadvertently erode the recognition of safety and inclusion - a critical trade-off for designers. \revt{Taken together, this pattern both reinforces prior HCI observations that overt nudges can provoke resistance rather than compliance~\cite{caraban201923}, and extends them by showing that, in the context of ableist microaggressions, such resistance can actually sharpen people's ability to differentiate harmful from neutral interactions. To our knowledge, this is the first empirical evidence in accessibility contexts that resisting biased AI suggestions can sharpen microaggression recognition, rather than simply leading to disengagement.}

\subsection{Mechanisms of Change: Inclusive Scaffolding and Self-Construction}

In contrast, the \textbf{Neutral-Directed} and \textbf{Self-Directed} conditions fostered a more balanced recognition by reinforcing inclusive norms. Neutral-directed participants perceived the coach's suggestions as helpful \textbf{"scaffolding"} that provided positive, actionable templates for respectful interaction. Both groups improved in affirming neutral interactions, observing or producing inclusive models of engagement.

Notably, however, these conditions did less to sharpen the \textbf{contrast} between neutral and ableist interactions compared to the biased condition. This suggests that \textbf{affirming the positive fosters recognition of inclusion but may not, by itself, clarify the boundaries of harm.} Furthermore, even inclusive scaffolding was sometimes seen as ``directive,'' highlighting the thin line between helpful guidance and unwelcome steering in supportive HCI systems.

\begin{revision}
    These dynamics suggest a complementary role for AI-mediated dialogue alongside disability-led education rather than as a replacement for it. Disability training and education work has emphasized the value of storytelling, role-play, and simulation exercises led by disabled people in building empathy and challenging ableist assumptions~\cite{french1992simulation,ioerger2019interventions,san2022use}. However, such activities are labor-intensive and often difficult to scale. Our findings point to one way interactive systems could extend this work: disability-led curricula can establish norms, share lived experience, and articulate why particular patterns of interaction are harmful, while AI-mediated dialogues provide low-stakes opportunities for repeated practice, feedback, and reflection between or after those sessions. This aligns with HCI research that calls for incorporating social factors and lived experience into accessible design processes~\cite{shinohara2018incorporating}, and highlights the importance of developing any such coaching systems in partnership with disabled communities. Our findings, therefore, complement and extend this disability-led training literature by providing empirical evidence for a concrete, AI-mediated practice layer that can operationalize these curricula between or after live sessions.
\end{revision}

\subsection{The Power of Active Engagement Over Passive Consumption}

Across nearly all measures, all three dialogue-based conditions outperformed the \textbf{Reading} condition. This shows the limits of passive exposure; simply presenting information about ableism did not reliably shift social judgments \revt{and, in our case, sometimes led to declines in recognition. Participants in the Reading condition became less likely to affirm neutral interactions as positive (Section~\ref{sec:results}), suggesting that reading about microaggressions without opportunities to practice responses may foster a more generally negative or skeptical stance toward social encounters without equivalently sharpening differentiation between harm and safety. This pattern resembles a kind of ``backfire effect,'' mirroring findings in diversity training research, where static instruction without behavioral practice can induce defensiveness or disconnects between abstract knowledge and situated application~\cite{chang2019mixed, forscher2019meta,lai2014reducing,lai2016reducing}}. In contrast, the dialogue conditions required participants to \textit{actively generate} responses in context. This act of construction was a far more effective mechanism for learning, consistent with theories of active learning \revt{that link more constructive and interactive forms of cognitive engagement to deeper learning outcomes~\cite{freeman2014active, chi2014icap}. Our findings echo calls in healthcare and education to move from knowledge-transfer interventions toward feedback-rich, interactive systems that help practitioners reflect on their language in situ~\cite{bascom2024designing, maqsood2025effect}. Our results reinforce these broader findings while extending them into the underexplored domain of disability microaggressions, showing experimentally that brief, AI-mediated conversational practice outperforms a reading-only module on short-term recognition and differentiation}. The \textbf{Self-Directed} group's emphasis on ``authenticity'' and ``contextual adjustment'' further shows that unguided, reflective practice can be a valuable learning experience in itself\revt{, aligning with HCI work on critical self-reflection and deliberation in human-AI collaboration~\cite{glinka2023critical, ma2025towards}. In the underexplored domain of disability microaggressions, our study therefore adds experimental support to these broader claims about the advantages of active, feedback-rich engagement over passive instruction}.

\subsection{Connecting to Broader HCI and Computational Social Science}

Our work contributes to several ongoing conversations in HCI and related fields:

\begin{itemize}
    \item \textbf{Framing and Nudging in HCI:} This study extends classic work on framing effects~\cite{tversky1981framing} \revt{and technology-mediated nudging~\cite{caraban201923,cockburn2020framing}} into the dynamic context of AI-mediated dialogue. We demonstrate that users are not passive recipients of AI ``nudges''\revt{: biased coaching was frequently resisted, and inclusive coaching was selectively adopted as scaffolding. This complicates design assumptions that interface defaults or suggestions will be quietly accepted, and aligns with recent HCI work that emphasizes the negotiated character of human-AI influence~\cite{ma2025towards,li2025actions}.}
    % ; they actively interpret, resist, or adopt them based on social and moral norms, complicating a simplistic view of nudging.
    \item \textbf{Human-AI Interaction as Social Simulation:} We present a novel paradigm for using LLMs not just as tools but as \textbf{social actors} within controlled experiments. \revt{This methodology complements interactive evaluation platforms such as SOTOPIA~\cite{zhou2023sotopia} by focusing on human outcomes rather than model scores: instead of asking whether an agent exhibits social intelligence, we ask how its prompts and framings reshape human judgments in situ. In doing so, we offer a methodological contribution in the form of a reusable experimental apparatus that other researchers can adapt to study different forms of social bias in interaction, alongside empirical evidence about AI-mediated bias interventions.}
    % This methodology offers a scalable way to study sensitive social phenomena with greater ecological validity than static surveys.
    \item \textbf{Microaggression Research:} By demonstrating that normative and affective judgments are distinct skills \revt{that can move differently across conditions}, our work \revt{complements psychological accounts of microaggressions as both everyday norm violations and sources of cumulative harm~\cite{sue2007racial,nadal2018microaggressions}. It also extends recent vignette-based HCI work on ableism~\cite{timmons2024ableism} by embedding scenarios in interactive dialogue and showing how different coaching framings modulate recognition trajectories over time.}
    % suggests that interventions should target both analytical recognition of rules and the empathetic understanding of why microaggressions cause harm.
        \item \textbf{Complementing Disability-Led Education:} 
        % We position this work as an \emph{empirical contribution} evaluating a new tool for accessibility education~\cite{wobbrock2016research}. Crucially, AI-mediated dialogue is not intended to replace training led by disabled people, which provides essential lived experience and authority. Instead, it offers a scalable ``sandbox'' for \textit{practice}. While disability-led workshops are resource-intensive and often limited to passive listening, AI agents can serve as tireless role-play partners, allowing learners to make mistakes and experience ``critical friction'' in a private setting before engaging in real-world interactions.
        We do not position AI-mediated dialogue as a replacement for training led by disabled people, which provides essential lived experience, context, and accountability. Instead, our findings suggest a complementary role: disability-led workshops and courses can establish norms and share lived experiences, while AI agents offer a scalable ``sandbox'' for practice, where learners can rehearse language, make mistakes, and encounter critical friction in a low-stakes environment between or after those sessions. In terms of HCI contribution types~\cite{wobbrock2016research}, this work provides empirical evidence about how different coaching directions shape bias recognition, a methodological contribution in the form of an AI-mediated dialogue platform and vignette corpus for studying ableism in situ, and design-oriented implications for socially aware, anti-ableist AI systems.

\end{itemize}

This study demonstrates that AI-driven dialogue can be a potent medium for social learning. The findings indicate a more intricate reality, one where the path to recognizing bias is not singular. While inclusive models build positive schemas, the active struggle against biased models can forge a sharper, more critical understanding of social harm.

\section{Design Implications for Socially-Aware AI}

Our findings carry direct implications for the design of AI systems that mediate or influence human social interaction. As conversational AI becomes more prevalent, understanding how subtle prompts shape social judgment is crucial for creating responsible and effective technology.

\begin{itemize}
    \item \textbf{Guidance is Not Neutral.} The starkly different outcomes between our conditions show that there is no ``neutral'' way to design a social AI. System defaults, suggested phrasings, and conversational patterns inevitably frame interactions and influence users. Designers should, therefore, treat coaching prompts as value-laden design decisions and make those norms explicit rather than presenting them as purely objective or generic.
    
    \item \textbf{Prefer Scaffolding Over Prescription.} The \textbf{Neutral-\hspace{0pt}Directed} condition suggests that users benefit most when AI provides options and examples that they can adapt, rather than a single ``correct'' response. In practice, this means generating a small set of alternative suggestions or question stems, and allowing users to ignore or modify them, rather than forcing the conversation down one path.

    \item \textbf{Balance Sensitivity with Positivity.} \revt{The \textbf{Bias-Directed} condition increased sensitivity to harm but also produced a more globally negative evaluation pattern. To avoid this, systems intended for training should not only highlight problematic phrasings or microaggressions but also immediately offer constructive, bias-aware alternatives. This pairing can support both vigilance and affirmation of safe, respectful interactions.}
    % The trade-off seen in the \textbf{Bias-Directed} condition is a critical lesson. An effective system must cultivate both vigilance against harm and an appreciation for positive interactions. Interventions could be designed to explicitly model both: first identifying and rejecting a problematic interaction, then immediately practicing a constructive alternative.

    \item \textbf{Consider "Critical Friction" Carefully and with consent.} \revt{Our results suggest that resisting biased suggestions can prompt useful reflection, but biased prompts also carry risk. If designers introduce intentionally problematic examples (e.g., in a training mode), they should clearly label this mode, obtain consent, and provide context about why the example is inappropriate. Outside of such controlled settings, systems should avoid generating biased framings in the first place.}
    % While exposing users to negative examples can be a powerful learning tool, it carries ethical risks. If used, it should be done with transparency and consent, framing it as a critical thinking exercise. The goal is to provoke reflection, not to normalize or accidentally teach harmful behavior.
    
    \item \textbf{Integrate With Disability-Led Education.} Finally, AI-mediated dialogue should be deployed as a complement to, not a replacement for, disability-led workshops and curricula. Our findings suggest that such tools are well-suited as scalable ``practice spaces'' where learners can rehearse language and reflect on their choices between live sessions, especially when co-designed with disabled people.
\end{itemize}

\section{Limitations and Future Work}

Our study's limitations provide important context for interpreting these findings and open several promising directions for future research.

First, our text-based dialogue, while enabling experimental control, lacks the multimodal richness of face-to-face interaction. Future work could explore these dynamics in more immersive environments, such as using \textbf{virtual reality or voice-based agents}, to incorporate non-verbal cues like tone and body language.

Second, our intervention was a single, brief exposure. While we demonstrated immediate shifts in judgment, we cannot assess the \textbf{durability} of these changes. Longitudinal studies are needed to track whether these effects persist over time and, crucially, if they \textbf{transfer to actual interpersonal behavior}.

Third, the artificial nature of the LLM-driven interaction may limit transferability, as modern LLMs still demonstrate significant limits in their social intelligence and theory of mind~\cite{sap2023neural}. \revt{Our platform is therefore best understood as an experimental apparatus that approximates everyday conversation rather than a full reproduction of real-world, multimodal interactions.} Future studies could use this platform in a ``Wizard of Oz'' setup to compare the effects of AI-guidance versus human-guidance in mediating conversations.

\begin{revision}
    Fourth, our sample demographics reflect specific limitations regarding generalizability. The majority of our participants (71\%) were aged 18--44, and the study was conducted exclusively in English. As perceptions of ableism and conversational norms often vary across generations and linguistic cultures, future work should deliberately recruit older adults and non-native English speakers to investigate how these factors shape the recognition of microaggressions and to test whether dialogue-based interventions and coaching directions produce similar effects in other contexts.
\end{revision}

Finally, the experimental paradigm itself is highly generalizable and can serve as a powerful tool for studying other forms of social bias beyond ableism. The core components, a simulated dialogue with an AI character representing a member of a marginalized group and an AI coach modeling biased or inclusive framings, can be readily adapted. For instance, researchers could study:
\begin{itemize}
    \item \textbf{Racial Microaggressions:} By simulating a job interview or a casual social encounter where biased nudges might suggest comments like, ``But where are you \textit{really} from?'' or play into stereotypes about competence or background.
    \item \textbf{Gender Bias:} By simulating a professional meeting where a biased coach suggests a participant interrupt a female colleague (``hepeating'') or question her technical expertise, allowing for the study of how people navigate and resist sexist framings in real-time.
    \item \textbf{Ageism:} By modeling workplace interactions where a younger manager receives biased prompts suggesting they patronize an older employee's technical skills or make assumptions about their career ambitions.
\end{itemize}
This framework serves as an important contribution to Computational Social Science by providing a novel methodology to empirically study bias. It creates a testbed for evaluating the societal impact and persuasive power of large language models, operationalizing key questions in AI Ethics about algorithmic influence and responsible AI design.

\section*{Conclusion}

This study demonstrates that the design of AI-mediated dialogue not only reflects social norms, but actively shapes them. Our findings reveal that while inclusive AI coaching provides helpful scaffolding for positive interaction, it was active resistance to biased nudges that most sharply honed the participants' ability to differentiate harm. This presents a fundamental tension for intervention design: the path to increased critical awareness may not be the same as the path to reinforcing positive social engagement.

Our primary contribution is twofold: an empirical demonstration of the nuanced, often contradictory effects of AI-driven social nudging and a generalizable experimental framework for studying these dynamics. This framework offers a robust methodology for HCI and Computational Social Science to move beyond static surveys and investigate bias recognition in a more ecologically valid, interactive context.

As we move into an era where conversational AI will increasingly mediate our social lives, our responsibility as designers extends beyond creating systems that are merely efficient or engaging. We must design systems that foster critical thought, respect human agency, and are intentionally built to support more inclusive and equitable human relationships. The central challenge is not simply to build AI that understands social norms but to build AI that helps us, as humans, understand them and each other better.

\begin{acks}
    We sincerely thank Christine Mendoza, Matildelis Medina, and Carrie Danny for generously sharing their time and expertise to strengthen the vignette materials. We also thank our participants for their time and contributions. This work was supported in part by Carnegie Mellon University's Presidential Postdoctoral Fellowship (awarded to the first author).
\end{acks}

\bibliographystyle{ACM-Reference-Format}
\bibliography{sample-base}

\end{document}